\documentclass[10pt,english,aps,manuscript,preprint, prb,floatfix,superscriptaddress,showpacs,amsmath,amssymb,reprint]{revtex4-1}
\usepackage[T1]{fontenc}
\usepackage[latin9]{inputenc}
\usepackage{color}
\usepackage{amssymb}
\usepackage{graphicx}
\usepackage{esint}
\usepackage[normalem]{ulem}
\usepackage{amsmath} 
\makeatletter

\@ifundefined{textcolor}{}
{%
 \definecolor{BLACK}{gray}{0}
 \definecolor{WHITE}{gray}{1}
 \definecolor{RED}{rgb}{1,0,0}
 \definecolor{GREEN}{rgb}{0,1,0}
 \definecolor{BLUE}{rgb}{0,0,1}
 \definecolor{CYAN}{cmyk}{1,0,0,0}
 \definecolor{MAGENTA}{cmyk}{0,1,0,0}
 \definecolor{YELLOW}{cmyk}{0,0,1,0}
}

\PassOptionsToPackage{caption=false}{subfig}

\makeatother

\usepackage{babel}
\begin{document}

\title{Magnification of signatures of topological phase
transition by quantum zero point motion}

\author{Pedro L. e S. Lopes}

\affiliation{Instituto de F\'{i}sica Gleb Wataghin, Universidade Estadual de Campinas,
Campinas, SP 13083-970, Brazil}

\author{Pouyan Ghaemi}

\affiliation{Physics Department, City College of the City University of New York, New York, NY 10031}
\begin{abstract}
 We show that the zero-point motion of a vortex in superconducting doped topological insulators leads to significant changes in 
 the electronic spectrum at the topological phase transition in this system. This topological phase transition is tuned by the doping level and the 
 corresponding effects are manifest in the density of states at energies which are of the order of the vortex fluctuation frequency. While the electronic energy gap
 in the spectrum generated by a stationary vortex is but a small fraction of the bulk superconducting gap, the vortex fluctuation frequency may be much larger. As a result, this 
 quantum zero-point motion can induce a discontinuous change in the spectral features of the system at the topological vortex phase transition to energies which are well 
 within the resolution of scanning  tunneling microscopy. This discontinuous change is exclusive to superconducting systems in which we have a topological phase transition. 
 Moreover, the phenomena studied in this work present novel effects of Magnus forces on the vortex spectrum which are not present in the ordinary 
 s-wave superconductors. Finally, we demonstrate explicitly that the vortex in this system is equivalent to a Kitaev chain. This allows for the mapping of the vortex
 fluctuating scenario in three dimensions into similar one dimensional situations in which one may search for other novel signatures of topological phase transitions.
\end{abstract}

\maketitle
\section{Introduction}Topologically distinct phases which cannot be classified by the classical Landau paradigm comprise some of the most recently discovered 
states of matter\cite{Wen,Levin,Burnell}. An important signature of these topological phases is the appearance of novel, low-energy, robust, edge states; one such state is the so-called 
Majorana bound state at the edges of topological superconductors \cite{Majorana}. As ubiquitous signatures, the detection of these neutral fermions has been the main trend in the 
characterization of particle-hole symmetric topological phases. Although evidences of Majorana fermion physics have been identified in tunneling \cite{delft} and scanning tunneling microscopy (STM) 
measurements\cite{andrey}, the interpretation of their signatures is controversial in many cases, as the imprints from the topological regime are often mixed with signals from disorder
and extra undesired quasiparticles. 

While the aforementioned gapless edge states act as a signature of topologically non-trivial regimes, 
the signatures of the \textit{transition} from a topologically trivial to a topological phase present themselves in the bulk by the closing and re-opening of the excitation energy 
gap\cite{Bernevig,Jay,ghaemisarang}. In many of the proposed systems which can be tuned through a topological phase transition (TPT), the excitation gap is very small 
compared with experimental resolutions and cannot be probed directly.

In this work, we show that quantum fluctuations can \textit{shift the spectral weight} in the density of states of a given system before and after a TPT to further 
separated energies and, as a result, magnify the change of the spectrum resulting from this process. This situation will be relevant as long as the sample's temperature is below
$\hbar\omega_0/k_B$, where $\omega_0$ is the pinning frequency and $k_B$ is Boltzmann's constant. We discuss this effect in the context of the chemical potential induced topological phase 
transition in the vortices of superconducting doped topological insulators. In this particular situation, we also demonstrate how the effects of Magnus forces on the vortex dynamics\cite{Ao} 
have a novel signature in the spectral change at this TPT, exposing the pumping of vortex modes responsible for the phase transition, as described below. Our results are general, however, 
and can be extended to other types of topological phase transitions. To demonstrate this, we present a way to map the 3D situation into a 1D setting in terms of wire networks which may be 
used to probe for the topological phase transition of actual Kitaev chains, Su-Schrieffer-Heeger chains and other unidimensional topological chains.

To understand how quantum fluctuations affect vortices in superconducting doped TIs we start by discussing vortex dynamics in regular superconductors (SCs).
This physics has been widely studied\cite{vortexmotion,Bardeen} and, given the natural length scale of vortices, their different properties might display both classical 
and quantum phenomena. Within the BCS theory of superconductivity, an stationary vortex affects the spectrum of the superconductor by generating in-gap modes localized around and along
the vortex core\cite{DeGennes}. The energy of these discrete bound states, known as  Caroli-de Gennes-Matricon (CdG) modes, is given by $\epsilon_l=\frac{\Delta^2}{\mu}\left(l+\frac{1}{2}\right)$ where $\Delta$ is the size of 
the bulk SC gap, $\mu$ is the fermi energy and $l$ is an integer. The signatures of these in-gap states have been experimentally observed by STM measurements\cite{DeGennesExp,Suderow}. In practice, however, 
even though the spatial resolution of STM is well within the size of the vortex modes\cite{STM}, given the small size of their so-called mini-gap, $\delta\equiv\frac{\Delta^2}{\mu}$, the energy of each single mode is  
hard to be resolved and usually multiple modes are observed together\cite{DeGennesExp}. 

It is well known that the  pinning of vortices is necessary for the stability of type-II SCs. The discussion above would be the final status of the problem for pinned vortices, were they 
absolutely static. Although a pinned vorticex has a fixed position at the sample, even at the lowest temperatures, their quantum zero point motion cannot be ignored. Interestingly, it was shown 
that such quantum fluctuations affect the quasiparticle spectrum, moving part of the spectral weights of the in-gap vortex modes to the frequencies associated with vortex fluctuations\cite{Bartosch,Nikolic,nikba}. 
We then contend that exploiting this ubiquitous quantum mechanical phenomenon to probe for TPTs is a promising idea, leading to novel signatures of these transitions.

To test this approach, superconducting doped TIs arise as the most natural test ground. The discovery of superconductivity in doped TIs triggered several studies, particularly because of 
the suggestions that doped TIs might realize topological superconductivity\cite{SCTI1,SCTI2,SCTI3,SCTI,SCTIFu}. Theoretical studies of superconductivity in the surface states of TIs 
started even before the experimental realization of bulk superconductivity in doped TIs, when it was shown that, theoretically, if superconductivity is induced in their helical surface 
states, vortex modes will include a zero-energy Majorana bound state\cite{fumajorana}. In the context of bulk superconducting doped TIs, it was later shown that 
the Majorana mode at the ends of a vortex line persist up to a critical value of doping in these systems as well\cite{pavan,ghaemitaylor,ghaemigilbert}. At this critical doping level, 
the two Majorana modes at the ends of the vortex hybridize and become gapped. The presence or absence of Majorana modes at the end of the vortex line contrast the two topologically 
distinct phases. In fact, the vortex in doped superconducting TIs becomes effectively equivalent to a Kitaev chain, one of the pioneering theoretical models to realize topological 
phases and phase transitions with Majorana edge states\cite{kitaev} (check also Section \ref{sec:1Dwire} of the present work).

As desired, the signature of this TPT also shows up in the spectrum of the states extended along the vortex. The original mechanism lies in the CdG modes. The important property 
of these states is that they are gapped by the small energy scale of the mentioned mini-gap. This energy protects the surface Majorana zero modes, confining 
them to the surface of the sample. Because of strong spin-orbit coupling and the resulting band inversion of TIs\cite{tiband}, the Fermi surface here has non-trivial topological properties which 
show up as a non-zero Berry connection. The CdG modes then inherit this Berry phase as a modification to their energy spectrum, which also separates in two sets due to the existence of two degenerate 
TI Fermi surfaces, which becomes $E_{l}^{\pm}=\frac{\Delta^{2}}{E_{F}}\left(l\mp\frac{1}{2}\pm\frac{\Phi_{b}\left(\mu\right)}{2\pi}\right)$. Here $\Phi_b$ is the Berry phase around the curve on the Fermi surface 
defined by setting the wave-vector along the vortex line equal to zero. In this case, when $\Phi_b=\pi$, $E^{\pm}_0=0$  and the zero energy surface Majorana modes at the ends of the vortex can  
merge through the gapless $l=0$ mode which is now extended along the vortex. The richness introduced by spin-orbit coupling and topology in this system leads to the signatures that we demonstrate.

For the physical picture of a fluctuating vortex to be reasonable, its position and cross-section structure must be well defined. Testing with some real numbers, Copper doped $Bi_2Se_3$ was 
the first topological insulator found to become superconducting upon doping, at 3.8 $K$\cite{SCTI1}. One must spatially resolve the local density of states (LDOS) at the vortex core which, 
as we demonstrate, comes from the $l=0$ and $l=1$ CdG modes. Their maxima lie at $r=0$ and are separated from the next closest mode (with $l=-1$) by the Fermi wavevector scale $r=1/k_F\approx 10 \mbox{\AA}$, which is
well within the resolution of STM. 
Regarding the energy scales, the change of spectrum at the TPT happens at the mini-gap energy scale $\delta \approx 5\times 10^{-3}\  K$. This is very small compared to the spectral 
resolution of STM which is of the order of the measurement temperature ($3k_BT$)\cite{STM}. It is then clear that an important obstacle to the verification of topological phase 
transitions in this system by STM  is the small excitation gap. Overcoming this energy scale problem is the main role of the vortex position fluctuation we analyze.

The paper is organized as follows. In Section \ref{sec:model} we explain the model in which we base our calculations. In Section \ref{sec:selfE} we follow Ref.\cite{Bartosch}, showing how the vortex fluctuations induce a 
self-energy correction which redistributes the peak weights in the LDOS for our specific model. This affects directly the tunneling conductance measured in STM experiments and in Section \ref{sec:analy} 
we demonstrate what are the novel consequences of this phenomenon for the vortex TPT in doped TIs. We believe that the approach we describe in the bulk of our paper is generalizable to other situations and we
dedicate section \ref{sec:1Dwire} to stipulate how to translate the ideas from the 3D context to 1D situations concerning Kitaev chains or other linear or quasi-linear topological phases. We 
conclude in Section \ref{sec:conc}. As computations are a bit involved, we avoid displaying them throughout our narrative as much as we can. We refer the reader to the appendices, where details are 
displayed thoroughly, whenever necessary.

\section{Fluctuating vortex model \label{sec:model}}

Superconductivity and the vortex quantum phase transition (VQPT) in doped topological insulators may be understood in the weak pairing limit ($\xi k_{F}\gg 1$, where $\xi$ is the SC coherence length)\cite{pavan}.
In this regime, a gradient expansion can be deployed to study the effects of the fluctuating vortex position in the low-energy spectrum\cite{Bartosch}.

We start with an action of the form $\mathcal{S}=\mathcal{S}_{BdG}+\mathcal{S}_{eff}^{vortex}$.
The first term is a Bogoliubov-de Gennes (BdG) action for the superconducting doped TI,
\begin{equation}
 \mathcal{S}_{BdG}=\frac{1}{2}\int d^{2}rd\tau\Psi^{\dagger}\left(\partial_{\tau}+H_{BdG}\right)\Psi \label{eq:SBdG}
\end{equation}
where 

\begin{eqnarray}
H_{BdG} & = & \left[\begin{array}{cc}
H_{TI}-\mu & \Delta\left(\mathbf{r}-\mathbf{R}\left(\tau\right)\right)\\
\Delta^{\dagger}\left(\mathbf{r}-\mathbf{R}\left(\tau\right)\right) & -H_{TI}+\mu
\end{array}\right]. \label{eq:BdGH}
\end{eqnarray}
Here $\mu$ is the chemical potential and the effective low energy 3D TI Hamiltonian is given by 
\begin{equation}
H_{TI}=-iv_D\tau_x \boldsymbol{s}.\boldsymbol{\nabla}+\tau_z \left( m+\epsilon\nabla^2 \right), 
\end{equation}
with Nambu-spinor $\Psi =\left(\psi,\, i s_{y}\psi^{\dagger}\right)^{T}$ and where $\psi=\left(\psi_{A\uparrow},\,\psi_{A\downarrow},\,\psi_{B\uparrow},\,\psi_{B\downarrow}\right)$. $A,\, B$ 
are orbital indices and $\tau_i$ and $s_i$ Pauli matrices act on orbital and spin Hilbert spaces, respectively. The superconducting pairing $\Delta\left(\mathbf{r}-\mathbf{R}\left(\tau\right)\right)$ contains
a vortex profile centered at a fluctuating position $\mathbf{R}\left(\tau\right)$
whose dynamics is governed by\cite{Bartosch}

\begin{eqnarray}
\mathcal{S}_{eff}^{vortex} & = & \frac{m_{v}}{2}\int\frac{d\omega}{2\pi}\mathbf{R}^{\dagger}\left(i\omega\right)\left(\begin{array}{cc}
\omega^{2}+\omega_{0}^{2} & \omega_{c}\omega\\
-\omega_{c}\omega & \omega^{2}+\omega_{0}^{2}
\end{array}\right)\mathbf{R}\left(i\omega\right).\nonumber \\
\label{eq:Svrtx}
\end{eqnarray}
Physically, the action (\ref{eq:Svrtx}) describes a particle of mass $m_{v}$ oscillating in an harmonic trap of frequency $\omega_{0}$ which depends on the properties of the 
trapping potential \cite{Bartosch}. This oscillator frequency dictates the qualitative features of the energy peak distribution of the LDOS. Finally, $\omega_{c}$ corresponds to a Magnus force acting on the vortex.
The frequency $\omega_{c}$ will be shown to play an essential role, introducing an energy scale for the chemical potential in which we have distinguished signatures of the VQPT in the system's LDOS.

\begin{figure}[t!]

\includegraphics[scale=0.12]{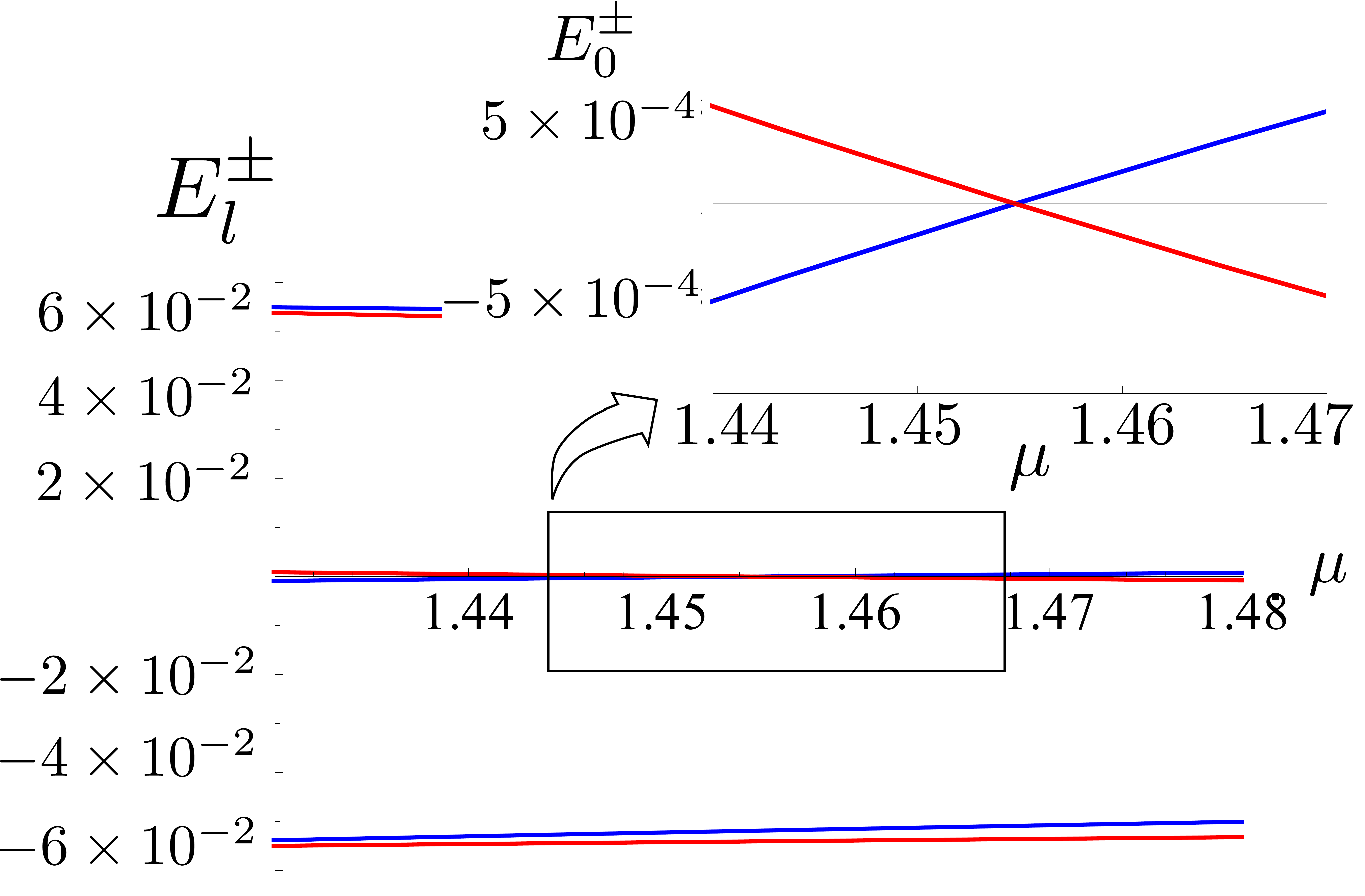}

\protect\caption{CdG vortex modes spectrum of the model from ref.\cite{pavan}. The parameters used
in the calculation are described in Figure \ref{fig:States} in the appendix.
We show the energies $E_{l}^{+}\left(\mu\right)$ (blue) and $E_{l}^{-}\left(\mu\right)$
(red) for $l=0,\pm1$ at chemical potentials close to $\mu_{c}\approx1.455$. The inset
displays the details for $l=0$. \label{fig:Solutions}}

\end{figure}

To capture the coupling between electronic excitations and vortex fluctuations, we expand the superconducting pairing around the
vortex rest position $\Delta\left(\mathbf{r}-\mathbf{R}\left(\tau\right)\right)\approx\Delta\left(\mathbf{r}\right)-\partial_{\mathbf{r}}\Delta\left(\mathbf{r}\right)\cdot\mathbf{R}\left(\tau\right)$. 
This approximation is valid at weak-coupling \cite{Bartosch}, which is also the regime of validity of Hamiltonian \eqref{eq:BdGH}. Within this formalism, the full problem is described by a 
perturbative action $\mathcal{S}=\mathcal{S}_{0}+\mathcal{S}_{eff}^{vortex}+\mathcal{S}_{int}.$ $\mathcal{S}_{0}$ is given by (\ref{eq:SBdG}) with the BdG Hamiltonian in the stationary vortex limit,
 $\mathbf{R}\left(\tau\right)=0$ (explicitly given in \eqref{eq:HBdG0}). The interaction term is given by
\begin{equation}
\mathcal{S}_{int}=-\int d^{2}rd\tau\mathbf{R}\left(\tau\right)\cdot\Psi^{\dagger}\left(\begin{array}{cc}
0 & \partial_{\mathbf{r}}\Delta\\
\partial_{\mathbf{r}}\Delta^{\dagger} & 0
\end{array}\right)\Psi . \label{eq:vrtxint}
\end{equation}
The interaction between vortex modes and the fluctuations in the vortex position leads to a self-energy correction to the energy of the CdG modes.

\section{Perturbed LDOS \label{sec:selfE}} Assuming a singlet intra-orbital pairing for doped TIs, the VQPT was found originally by an exact 
diagonalization of lattice toy models and a semi-classical study of the BdG mean-field Hamiltonian \cite{pavan}, as well as numerically solving the self-consistent 
BdG equations\cite{ghaemigilbert}. In order to study the effects of vortex fluctuations on the LDOS, it is convenient to use a basis which diagonalizes the Hamiltonian at the limit
of a static vortex. Thus, we first present the VQPT by a novel real-space diagonalization of the BdG equation of Hamiltonian \eqref{eq:HBdG0} following the ideas from \cite{Gygi}. 
The details follow in Appendix \ref{app:appA}.We expand the Grassmann fields in terms of eigenvectors of the static-vortex BdG Hamiltonian $H_{BdG}^{0}$ as $\Psi=\sum_{q=1}^{8}\sum_{ln}\chi_{ln}^{q}\left(\mathbf{r}\right)\psi_{ln}^{q}\left(\tau\right)$.
The eight arising bands obey $H_{BdG}^{0}\chi_{ln}^{q}\left(\mathbf{r}\right)=E_{ln}^{q}\chi_{ln}^{q}\left(\mathbf{r}\right)$ where $l$ and $n$ labels conserved quantum numbers. Precisely, $l$ represents 
a generalized angular momentum $\tilde{L}_{z}=-i\partial_{\theta}-\frac{s_{z}+\rho_{z}}{2}$, which commutes with the Hamiltonian (see \cite{pavan} or Appendix \ref{app:appA}), while $n$ labels
the different eigenstates of the radial BdG equation at fixed $l$. At weak coupling, we further project into the two bands which cross the doubly degenerate Fermi surface of $H_{TI}$. 
Labeling these states by $\sigma\equiv\pm$, we have, at low energies, $\Psi\approx\sum_{ln}\chi_{ln}^{+}\left(\mathbf{r}\right)\psi_{ln}^{+}\left(\tau\right)+\chi_{ln}^{-}\left(\mathbf{r}\right)\psi_{ln}^{-}\left(\tau\right)$ with

\begin{widetext}
\begin{eqnarray}
\chi_{ln}^{+}\left(\mathbf{r}\right)=\frac{1}{\sqrt{2\pi}}\int dk\left(\begin{array}{c} \label{eq:chip}
\frac{c_{lk}^{n}}{\sqrt{\mathcal{N}_{k}^{+}}}\left(\begin{array}{c}
e^{-i\left(l-1\right)\theta}kJ_{l-1}\left(kr\right)\\ 0
\\ 0
\\
e^{-il\theta}\left(m_{k}-\sqrt{m_{k}^{2}+k^{2}}\right)J_{l}\left(kr\right)
\end{array}\right)\\
\frac{d_{lk}^{n}}{\sqrt{\mathcal{N}_{k}^{-}}}\left(\begin{array}{c}
e^{-il\theta}\left(m_{k}+\sqrt{m_{k}^{2}+k^{2}}\right)J_{l}\left(kr\right)\\ 0 
\\ 0
\\
ke^{-i\left(l+1\right)\theta}J_{l+1}\left(kr\right)
\end{array}\right)
\end{array}\right)=\left(\begin{array}{c}
\mathbf{u}_{ln}^{+}\left(\mathbf{r}\right)\\
\mathbf{v}_{ln}^{+}\left(\mathbf{r}\right)
\end{array}\right)\\
\chi_{ln}^{-}\left(\mathbf{r}\right)	=	\frac{1}{\sqrt{2\pi}}\int dk\left(\begin{array}{c} \label{eq:chim}
\frac{\bar{c}_{lk}^{n}}{\sqrt{\mathcal{N}_{k}^{-}}}\left(\begin{array}{c}
0 \\
e^{-il\theta}\left(m_{k}+\sqrt{m_{k}^{2}+k^{2}}\right)J_{l}\left(kr\right)\\
e^{-i\left(l-1\right)\theta}kJ_{l-1}\left(kr\right)\\
0 \\
\end{array}\right)\\
\frac{\bar{d}_{lk}^{n}}{\sqrt{\mathcal{N}_{k}^{+}}}\left(\begin{array}{c}
0 \\
e^{-i\left(l+1\right)\theta}kJ_{l+1}\left(kr\right)\\
e^{-il\theta}\left(m_{k}-\sqrt{m_{k}^{2}+k^{2}}\right)J_{l}\left(kr\right)\\
0 \\
\end{array}\right)
\end{array}\right)=\left(\begin{array}{c}
\mathbf{u}_{ln}^{-}\left(\mathbf{r}\right)\\
\mathbf{v}_{ln}^{-}\left(\mathbf{r}\right)
\end{array}\right).
\end{eqnarray}
\end{widetext} 

The numerical diagonalization may be done replacing the infinite system with a disk of finite radius $R$ with a profile $\Delta_{0}\left(r\right)=\Delta_{0}\tanh\left(r/\xi\right)$ for the vortex 
and solving the secular equation for the Fourier-Bessel coefficients $c_{lk}^{n},\,d_{lk}^{n},\,\bar{c}_{lk}^{n}$ and $\bar{d}_{lk}^{n}$ (details follow in Appendix \ref{app:appA} and references therein.)

To study the VQPT we consider the lowest energy vortex modes. These are the CdG modes and allow fixing the label $n\rightarrow n_{CdG}$, which we drop. 
The two sectors (labeled by $\sigma=\pm$) are connected by particle-hole (PH) conjugation $\mathcal{C}=\rho_y {s}_y \mathcal{K}$ operator ($\mathcal{K}$ is the complex 
conjugation operator) as $\mathcal{C}\chi_{l}^{+}=\chi_{-l}^{-}$. The energies of the CdG vortex modes in this case are the expected\cite{pavan} 

\begin{equation}
E_{l}^{\pm}=\frac{\Delta^{2}}{E_{F}}\left(l\mp\frac{1}{2}\pm\frac{\Phi_{b}\left(\mu\right)}{2\pi}\right), \label{eq:thespectrum}
\end{equation}
so that $E^{+}_l=-E^{-}_{-l}$. Here $\Phi_{b}\left(\mu\right)$ is the Berry phase calculated around the Fermi surface on the curve with zero wavevector along the vortex\cite{pavan}. As the chemical 
potential increases, the Fermi surface enlarges and $\Phi_{b}\left(\mu\right)$  varies from $0$ to $2\pi$, defining
a critical chemical potential such that $\Phi_{b}\left(\mu_{C}\right)=\pi$. Our results for the energies of the CdG modes, which are presented in 
Fig. \ref{fig:Solutions}, are consistent with the previous study of the phase transition in Refs. \cite{pavan} and \cite{ghaemigilbert}.

In terms of the CdG eigenstates,  equation (\ref{eq:vrtxint}) is written

\begin{eqnarray}
\mathcal{S}_{int} & = & -\sum_{l,l^{'},\sigma}\int d\tau\bar{\psi}_{l}^{\sigma}\left(\tau\right)\psi_{l^{'}}^{\sigma}\left(\tau\right)\mathbf{R}\left(\tau\right)\cdot\mathbf{M}_{l,l^{'}}^{\sigma},\nonumber \\ 
\end{eqnarray}
where
\begin{equation}
\mathbf{M}_{l,l^{'}}^{\sigma} = \int d^{2}r\chi_{l}^{\sigma}\left(\mathbf{r}\right)^{\dagger}\left(\begin{array}{cc}
 0 & \partial_{\mathbf{r}}\Delta\\
 \partial_{\mathbf{r}}\Delta^{\dagger} & 0
 \end{array}\right)\chi_{l^{'}}^{\sigma}\left(\mathbf{r}\right). 
\end{equation}
Vortex fluctuations then generate the following self-energy for CdG vortex modes which we calculate using the GW approximation\cite{GW} (details follow in Appendix \ref{app:appB}), 

\begin{equation}
\Sigma_{l}^{\sigma}\left(i\tilde{\omega}\right) = \sum_{l^{'},\alpha=\pm}\frac{A_{l;l^{'}}^{\alpha;\sigma}}{\left(i\tilde{\omega}-\left(sgn\left(\Xi_{l^{'}}^{\alpha;\sigma}\right)\omega_{v}\right)-\Xi_{l^{'}}^{\alpha;\sigma}\right)} \label{eq:selfie}
\end{equation}
Here $A_{l;l^{'}}^{\alpha;\sigma}\equiv\frac{\left|M_{l,l^{'}}^{\alpha;\sigma}\right|^{2}}{m_{v}\omega_{v}}$
are reduced matrix elements with $M_{l,l^{'}}^{\alpha;\sigma}=\frac{1}{2}\left(M_{x}+\alpha iM_{y}\right)_{l,l^{'}}^{\sigma}$
and $\Xi_{l^{'}}^{\alpha;\sigma}\equiv E_{l^{'}}^{\sigma}+\alpha\omega_{c}/2.$
For unit vorticity, angular momentum conservation implies that $l$
is connected only to $l^{'}=l+\alpha1$ by such interactions. The energy scale
introduced by $\omega_{v}\equiv\sqrt{\omega_{0}^{2}+\omega_{c}^{2}/4}$ (
and dominated by $\omega_{0}$ as aforementioned), represents a ``magneto-plasma''
frequency in an Einstein model\cite{Bartosch}. In Appendix \ref{app:appB}, we present closed formulas for these matrix elements.

One finally needs to evaluate the LDOS,
\begin{equation}
\rho\left(\mathbf{r},\omega\right)=\sum_{m,\sigma,l}\left|\left\langle \epsilon_{m}\left|\psi_{\sigma,l}^{\dagger}\left(\mathbf{r}\right)\right|N_{0}\right\rangle \right|^{2}\delta\left(\omega-\epsilon_{m}\right),
\end{equation}
where $\left|N_{0}\right\rangle $ is a $N_{0}$-particle ground-state,  $\left|\epsilon_{m}\right\rangle $ is an $\left(N_{0}+1\right)$-particle
excited state (with generic quantum numbers $m$) and $\psi_{\sigma,l}^{\dagger}\left(\mathbf{r}\right)$ is an electronic state creation operator at level $l$ in sector $\sigma$. Using the vortex-modes eigenbasis, this can be written, taking into account
the effects of the vortex fluctuations in the self-energy, as
\begin{eqnarray}
\rho\left(\mathbf{r},\omega\right) & = & \sum_{\sigma=\pm}\rho_{\sigma}\left(\mathbf{r},\omega\right)\\
\rho_{\sigma}\left(\mathbf{r},\omega\right) & = & -\frac{1}{\pi}Im\sum_{l}\frac{\left|\mathbf{u}_{l}^{\sigma}\left(\mathbf{r}\right)\right|^{2}}{\omega-E_{l}^{\sigma}-\Sigma_{l}^{\sigma}+i\epsilon}.\\
& = & \sum_{l} \left|\mathbf{u}_{l}^{\sigma}\left(\mathbf{r}\right)\right|^{2}\delta\left(\omega-E_{l}^{\sigma}-\Sigma_{l}^{\sigma}\left(\omega\right)\right).\label{eq:Local}
\end{eqnarray}

Through the perturbative interaction, the energy density profile of CdG modes is modified with part of the spectral weight from
$\omega=E_{l}^{\sigma}$ being transfered to new ``satellite'' peaks
in the LDOS \cite{Bartosch}. Both the spectrum $E_{l}^{\sigma}$ and the profile of $\mathbf{u}_{l}^{\sigma}\left(\mathbf{r}\right)$
dramatically change the phenomenology described by (\ref{eq:Local}) when the parent metallic state of the superconductor comes from doped TIs, as compared with ordinary metals.

\section{Tunneling Conductance Analysis \label{sec:analy}}
The local tunneling conductance is found, at low temperatures, by convolving the LDOS \eqref{eq:Local} with the derivative of the Fermi distribution function as
\begin{equation}
G\left(\mathbf{r},\omega\right)=-\frac{G_{0}}{\rho_{0}}\int d\omega^{'}\rho\left(\mathbf{r},\omega+\omega^{'}\right)f^{'}\left(\omega^{'}\right).
\end{equation}
The normalization constant assumes an STM tip with constant DOS $\rho_{0}=m_{e}/2\pi$ (for a free 2D electron gas) with the corresponding tunneling conductance $G_{0}$, and $f\left(\omega\right)$ is the Fermi-Dirac distribution.
At very low temperatures, the tunneling conductance is equal to  the LDOS, however still smoothed by the finite temperature effects.

Given the atomic level resolution of STM, we can safely focus at the density of states at the vortex core $\mathbf{r}=0$.
As seen in \eqref{eq:chip} and \eqref{eq:chim}, the wavefunction components may be expanded in terms of Bessel functions. In particular, at $\mathbf{r}=0$, only Bessel functions of order zero have 
non-zero amplitude while all the other Bessel functions vanish. From our Fourier-Bessel expansion of the CdG modes above, only $l=0$ and, as a result of spin-orbit coupling, $l=1$ modes have finite contributions 
in  $\mathbf{u}_{l}^{\sigma}\left(\mathbf{r}\right)$ at the origin. 

The $l=0$ states have energies $E_{0}^{\sigma}=\sigma\frac{\Delta^{2}}{E_{F}}\left(-\frac{1}{2}+\frac{\phi\left(\mu\right)}{2\pi}\right)$.
These energy levels may be pumped from negative to positive values (and vice versa) by changing the chemical potential, evolving the Berry phase from $0$ to $2\pi$. This novel feature leads to a change of sign in the factors of $\omega_{v}$ in the self-energy given in (\ref{eq:selfie})
when $l^{'}=0$, which determine the energies of the satellite peaks. As a result, the TPT manifests itself by a discontinuous change in the density of states by energies of order $\omega_v$ to energies of order $-\omega_v$. 

Remarkably, the local spectrum at the vortex center breaks particle-hole symmetry. The origin of this lies in the spin-orbit coupling which, together with the BdG doubling, filtered only the states $l=0,1$ at this position, leaving 
out the $l=-1$ states. Naturally the full DOS is PH symmetric. These points will be considered again in Sec. \ref{sec:1Dwire} in the context of the effective theory for the vortex bound states
after integration in the radial and angular directions. 

Even more importantly, one notes that the Magnus force term associated with the vortex motion, whose amplitude is proportional to $\omega_c$, breaks the mirror symmetry which is connecting the PH 
sectors of the CdG modes. As a result, the discontinuous transition of energy of the CdG modes from the two $\sigma$ sectors does not happen 
simultaneously at the same value of doping for both cases. This is essential for the change in the LDOS to be seen in this context, as it provides an energy window over which the density of states at the 
energy of vortex oscillations is remarkably modified by the TPT. It is also important to note that, for other CdG modes (such as the mode with $l=-1$ whose maximum amplitude is at $1/k_F\approx 10 \mbox{\AA}$
away from the center of the vortex), the opposite transition will happen. Given the spatial resolution of STM, however, the different modes should be resolvable. 

To make the above claims regarding the peak-jumping less abstract, let us concretely analyse the relevant contributions to the self-energy. As discussed, from angular momentum conservation,
$M_{l,l^{'}}^{+;\sigma}=\delta_{l^{'},l+1}M_{l,l+1}^{+;\sigma}$ and from $A_{l;l^{'}}^{-;\sigma}=A_{l^{'};l}^{+;\sigma}$ we can read
the corresponding result for $\alpha=-$. These simplifications allow us to reduce the self-energy to just a couple of relevant pieces,
\begin{widetext}
\begin{eqnarray}
\Sigma_{0}^{\sigma}\left(\omega\right) & = & \frac{A_{0;1}^{+;\sigma}}{\left(\omega-sgn\left(\Xi_{1}^{+;\sigma}\right)\omega_{v}-E_{1}^{\sigma}-\omega_{c}/2\right)}+\frac{A_{-1;0}^{+;\sigma}}{\left(\omega-sgn\left(\Xi_{-1}^{-;\sigma}\right)\omega_{v}-E_{-1}^{\sigma}+\omega_{c}/2\right)} \label{eq:sigma0}
\end{eqnarray}
and
\begin{eqnarray}
\Sigma_{1}^{\sigma}\left(\omega\right) & = & \frac{A_{1;2}^{+;\sigma}}{\left(\omega-sgn\left(\Xi_{2}^{+;\sigma}\right)\omega_{v}-E_{2}^{\sigma}-\omega_{c}/2\right)}+\frac{A_{0;1}^{+;\sigma}}{\left(\omega-sgn\left(\Xi_{0}^{-;\sigma}\right)\omega_{v}-E_{0}^{\sigma}+\omega_{c}/2\right)}.\label{eq:sigma1}
\end{eqnarray}
\end{widetext}
To find the positions of the peaks, one solves
\begin{equation}
 \omega-E_{l}^{\sigma}-\Sigma_{l}^{\sigma}\left(\omega\right)=0.
\end{equation}
The solutions are clearly sensitive to the sign of $\Xi_{l^{'}}^{\alpha;\sigma}\equiv E_{l^{'}}^{\sigma}+\alpha\omega_{c}/2.$  As we do not have an estimate for the actual strength of the 
Magnus effect, to be definite, we take $\omega_{c}=\eta\frac{\Delta^{2}}{\mu}$. It is a simple job to notice that  $sgn\left(\Xi_{1}^{+;\sigma}\right)=sgn\left(\Xi_{2}^{+;\sigma}\right)=+$ and $sgn\left(\Xi_{-1}^{-;\sigma}\right)=-$, 
for any value of the chemical potential. The sign of $\Xi_{0}^{-;\sigma}$, however, does depend on $\mu$. This allows one to define a value $\bar{\mu}_{\sigma}$
at which $sgn\left(\Xi_{0}^{-;\sigma}\right)$ changes.  The structure of $\Sigma_{1}^{\sigma}\left(\omega\right)$ depends crucially on this. From the CdG spectrum \eqref{eq:thespectrum}, we set $\Xi_{0}^{-;\sigma}=0$ explicitly, finding
\begin{eqnarray}
-\sigma+\left(\sigma\frac{\phi\left(\bar{\mu}_{\sigma}\right)}{\pi}-\eta\right) & = & 0\nonumber \\
\Rightarrow\frac{\phi\left(\bar{\mu}_{\sigma}\right)}{\pi} & = & 1+\eta\sigma,
\end{eqnarray}
where $\phi\left(\mu\right)$ is the Berry phase. As this phase grows monotonically from $0$ to $2\pi$, it is clear that the sector $\sigma=+$ has
a sign change at values of $\mu$ larger than those of the $\sigma=-$ sector, as long as $\eta\neq0$. In sum, this determines when each set of peaks will jump as 
function of $\mu$. In Appendix \ref{app:appC} we explore this further, also showing analytically that, at $\textbf{r}=0$, only the leftmost satellite peak from $l=1$ will jump due to this sign change.

\begin{figure}[t]
\includegraphics[scale=0.60]{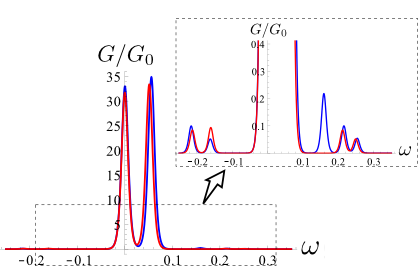}

\protect\caption{Tunneling conductance for $\mu<\bar{\mu}_{\pm}$,
and $\mu>\bar{\mu}_{\pm}$ in blue, and $\bar{\mu}_{-}<\mu<\bar{\mu}_{+}$, in red. The large central peaks correspond to $\omega \approx E_{l=0}^{\pm}$
and $\omega \approx E_{l=1}^{\pm}$ (the energies of the CdG modes for stationary vortex). The inset displays the effects of vortex
fluctuations. 
The smaller sattelite peaks appear at energies close to $\approx \pm \omega_v$.  Red curves correspond to $\mu<\bar{\mu}_{-}$ or $\mu>\bar{\mu}_{+}$.\label{fig:Peaks}}

\end{figure}

Figure \ref{fig:Peaks} displays our main results. It shows the differential
conductivity at the vortex center $G\left(\mathbf{r}=0,\omega\right)$ (more details on numerical parameters used here are given in Figure \ref{fig:States} in the appendix). Angular momentum conservation implies that each non-interacting energy 
level unfolds into a set of three peaks. 

We present the differential conductance for three ranges of chemical potential $\mu<\bar{\mu}_{\pm}$,
in blue, and $\bar{\mu}_{-}<\mu<\bar{\mu}_{+}$, in red, and $\mu>\bar{\mu}_{\pm}$ in blue again, which appears to be identical to $\mu<\bar{\mu}_{\pm}$. This happens because the separation
of the central peaks from $l=1$ is $E_{1}^{+}-E_{1}^{+}=\delta (-1/2+\phi(\mu)/2\pi)$, which cannot be resolved close to the phase transition (just as the peaks from $l=0$ cannot be resolved at this 
situation.) In this situation, having a finite $\omega_c$ is crucial to observe all peaks and the discontinuous effects of the topological phase transition. The pattern in the LDOS should be, for each $l$ and sector $\sigma$, of a 
large central peak located at $E_{l}^{\sigma}$ with the two partners offset approximately by $\pm\Omega_{l}^{\sigma}$ with $\Omega_{l}^{\sigma}=\sqrt{\left(\omega_{v}+\delta+\omega_{c}/2\right)^{2}+2A_{l}^{+;\sigma}}$.
In our case, a total of 12 peaks is expected for each value of the chemical potential (3 from $l=0$, another 3 from $l=1$ and twice this due to the two sectors), not all of them being resolvable due to thermal effects.
The large peaks closest to $\omega=0$ correspond to $\omega \approx E_{l=0,1}^{\sigma}$. The strength of the respective satellite peaks is suppressed by a $\xi^{-5}$ factor, where
$\xi$ is the coherence length \cite{Bartosch}. An inset displays the position of these peaks.

A remarkable behavior develops in the $l=1$ sattelite peaks (the rightmost small peaks at negative and positive frequencies). 
This is evidenced by the solitary blue peak at positive $\omega$. It corresponds to the contribution coming from $\omega=E_{1}^{-}-\Omega_{1}^{-}$, whose position jumps from this value
by approximately $2\omega_{v}$ as the chemical potential pumps the negative energy state at $E_{0}^{-}$ into positive energies after crossing $\bar{\mu}_{-}$. Similarly, when $\mu$ moves above $\bar{\mu}_{+}$, the
peak from $\omega=E_{1}^{+}+\Omega_{1}^{+}$ jumps by $-2\omega_{v}$. In appendix \ref{app:appC} we demonstrate that the approximate positions of the $l=1$ peaks can be determined analytically.

Concerning the magnitude of the Magnus effect,  if $\eta<1$, the effects from the Magnus force are sub-dominant to the CdG energy gap and the sensibility to which one needs to tune the (zero-temperature) chemical potential may again be beyond technical realization at the current time. If $\eta>1$, on the other hand, as the evolution of the Berry phase is from $0$ to $2\pi$, the critical chemical potentials $\bar{\mu}_\sigma$ may not be captured as one tune $\mu$ and one will be bound to the regime of $\bar{\mu}_{-}<\mu<\bar{\mu}_{+}$, which is similar to the standard s-wave case (except for the multiplicities of peaks and apparent breaking of the PH constraint.) As this seems to critically constrain the actual visualization of these effects in practice, we proceed now to consider some different situations in which one may actually control the energy difference between the  $l=1$ states for different $\sigma=\pm$  sectors. In this case, we will see that if  this energy difference can be made larger, even at 
$\eta=0$ one may be able to capture the closing and reopening of the energy gap from $l=0$.

\section{1D wire mapping  \label{sec:1Dwire}}

To conclude our considerations, we would like to speculate about the realization of similar signatures of TPTs by quantum motion in other systems. Here we demonstrate concretely the claim from 
\cite{pavan} stating that the vortex in superconducting doped TIs presents a topological phase transition equivalent to a Kitaev wire. We then proceed to showing that, more generally, the 
Hamiltonian projected at the vortex states corresponds to a set of wires (or a single multiband wire) inheriting a first neighbor mutual coupling from the vortex fluctuations in 3D. We then identify the important
ingredients necessary to realize the discussed phenomena in the context of 1D topological systems.

\subsubsection{Vortex Hamiltonian projection}
Start with Hamiltonian \eqref{eq:HBdG0} from the appendix, keeping the $z$-direction terms. We also keep the vortex fluctuations to first order in the gradient of the superconducting pairing. We have
\begin{eqnarray}
 H_{BdG}&=& H_{BdG}^{0}+\Gamma_{z} P_{z}-\Gamma_{0}\epsilon P_{z}^{2}\\ \nonumber
 &  & -\mathbf{R}\left(\tau\right)\cdot\left[\boldsymbol{\Lambda}\cdot\partial_{\mathbf{r}}\boldsymbol{\Delta}\left(\mathbf{r}\right)\right]\\
 &\equiv& H_{BdG}^{0}+H_{z}+V\left(\mathbf{r}\right).
\end{eqnarray}

We are going to project this into the lowest energy sectors $\chi_{ln}^{\pm}\left(\mathbf{r}\right)$ from \eqref{eq:chip} and \eqref{eq:chim}. At finite z, we have 
$\chi_{ln}^{\pm}\left(\mathbf{r}\right)\rightarrow\chi_{ln_{CdG}}^{\pm}\left(\mathbf{r}\right)f_{l}^{\pm}\left(z\right)$, choosing the CdG states with $n=n_{CdG}$. We 
will project the radial part the Hamiltonian to find out what Hamiltonian gives the equations of motion for $f_{l}^{\pm}$. Considering the $\pm$ sectors then we have
\begin{eqnarray}
 \tilde{H}_{ll^{'}} & = & Proj\left[H_{BdG}\right]_{ll^{'}}\\
 &=&\left(\begin{array}{cc}
E_{l}^{+} & 0\\
0 & E_{l}^{-}
\end{array}\right)\delta_{ll^{'}}\\&&+\left(\begin{array}{cc}
H_{zll^{'}}^{++} & H_{zll^{'}}^{+-}\\
H_{zll^{'}}^{-+} & H_{zll^{'}}^{--}
\end{array}\right)\\&&+\left(\begin{array}{cc}
V_{ll^{'}}^{++} & 0\\
0 & V_{ll^{'}}^{--}
\end{array}\right).
\end{eqnarray}

Notice that as $\left\langle \chi_{l}^{+}|\rho_{x}|\chi_{l}^{-}\right\rangle =\left\langle \chi_{l}^{+}|\rho_{y}|\chi_{l}^{-}\right\rangle =0$, the fluctuating vortex potential becomes 
diagonal with respect to the $\pm$ sectors. This result is the same as we had found in our considerations at vanishing $k_{z}$  and we already know what this term looks like,
\begin{eqnarray}
\left(\begin{array}{cc}
V_{ll^{'}}^{++} & 0\\
0 & V_{ll^{'}}^{--}
\end{array}\right)=\mathbf{R}\left(\tau\right)\cdot\mathbf{M}_{l,l^{'}}^{\sigma}
\end{eqnarray}
with
\begin{eqnarray}
\mathbf{M}_{l,l^{'}}^{+} & =& d^{2}r\left[\mathbf{u}_{l}^{+}\left(\mathbf{r}\right)^{\dagger}\partial_{\mathbf{r}}\Delta\mathbf{v}_{l^{'}}^{+}\left(\mathbf{r}\right)+\mathbf{v}_{l}^{+}\left(\mathbf{r}\right)^{\dagger}\partial_{\mathbf{r}}\Delta^{\dagger}\mathbf{u}_{l^{'}}^{+}\left(\mathbf{r}\right)\right] \nonumber \\ 
\\
\mathbf{M}_{l,l^{'}}^{-} & =& \int d^{2}r\left[\mathbf{v}_{l}^{+}\left(\mathbf{r}\right)^{\dagger}\partial_{\mathbf{r}}{\Delta}\mathbf{u}_{l^{'}}^{+}\left(\mathbf{r}\right)+\mathbf{u}_{l}^{+}\left(\mathbf{r}\right)^{\dagger}\partial_{\mathbf{r}}\Delta^{\dagger}\mathbf{v}_{l^{'}}^{+}\left(\mathbf{r}\right)\right].\nonumber \\ 
\end{eqnarray}
Due to the vortex structure in $\Delta$, we are only coupling $l$  to $l^{'}=l\pm1$.

Now we project $H_{z}$. To keep the notation short, we introduce $4\times 4$ Dirac matrices $\boldsymbol{\alpha}=\tau_{x}\boldsymbol{\sigma}$ and $\beta=\tau_{z}\sigma_{0}$ as in the appendix. It is easy to see that terms linear in $P_z$ contribute off-diagonal in the $\pm$ sectors while terms quadratic with $P_{z}^{2}$ contribute only diagonally. For these
diagonal terms, we develop couplings
\begin{eqnarray}
 \epsilon\rightarrow\begin{cases}
\epsilon_{l}^{+}=\epsilon\int d^{2}r\left[\mathbf{u}_{l}^{+}\left(\mathbf{r}\right)\right]^{*}\beta\mathbf{u}_{l}^{+}\left(\mathbf{r}\right)-\left[\mathbf{v}_{l}^{+}\left(\mathbf{r}\right)\right]^{*}\beta\mathbf{v}_{l}^{+}\left(\mathbf{r}\right)\\
\epsilon_{l}^{-}=\epsilon\int d^{2}r\left[\mathbf{u}_{l}^{-}\left(\mathbf{r}\right)\right]^{*}\beta\mathbf{u}_{l}^{-}\left(\mathbf{r}\right)-\left[\mathbf{v}_{l}^{-}\left(\mathbf{r}\right)\right]^{*}\beta\mathbf{v}_{l}^{-}\left(\mathbf{r}\right)
\end{cases}
\end{eqnarray}
Importantly, the sign of these couplings is the same and the angular integration enforces $l=l^{'}$. For the off-diagonal terms we develop the couplings
\begin{equation}
 \tilde{\Delta}_{l}=\int d^{2}r\left[\mathbf{u}_{l}^{+}\left(\mathbf{r}\right)\right]^{*}\alpha^{z}\mathbf{u}_{l}^{-}\left(\mathbf{r}\right)-\left[\mathbf{v}_{l}^{+}\left(\mathbf{r}\right)\right]^{*}\alpha^{z}\mathbf{v}_{l}^{-}\left(\mathbf{r}\right)
\end{equation}

\subsubsection{1D Wire network}
Adding up the matrix elements above gives the projected Hamiltonian
\begin{eqnarray}
 \tilde{H}_{ll^{'}}&=&\left(\begin{array}{cc}
E_{l}^{+}-\epsilon_{l}^{+}\partial_{z}^{2} & -i\tilde{\Delta}_{l}\partial_{z}\\
-i\tilde{\Delta}_{l}\partial_{z} & E_{l}^{-}+\epsilon_{l}^{-}\partial_{z}^{2}
\end{array}\right)\delta_{ll^{'}}\\&&+\left(\begin{array}{cc}
\mathbf{R}\left(\tau\right)\cdot\mathbf{M}_{l,l^{'}}^{+} & 0\\
0 & \mathbf{R}\left(\tau\right)\cdot\mathbf{M}_{l,l^{'}}^{-}
\end{array}\right).
\end{eqnarray}
For Hermiticity $\mathbf{M}_{l,l^{'}}^{\pm*}=\mathbf{M}_{l^{'},l}^{\pm}$. For the diagonal terms we may still use $E_{-l}^{-}=-E_{l}^{+}$ to write
\begin{eqnarray}
& &  \left(\begin{array}{cc}
E_{l}^{+}-\epsilon_{l}^{+}\partial_{z}^{2} & -i\tilde{\Delta}_{l}\partial_{z}\\
-i\tilde{\Delta}_{l}\partial_{z} & E_{l}^{-}+\epsilon_{l}^{-}\partial_{z}^{2}
\end{array}\right)\\
& = & \left(\begin{array}{cc}
E_{l}^{+}-\epsilon_{l}^{+}\partial_{z}^{2} & -i\tilde{\Delta}_{l}\partial_{z}\\
-i\tilde{\Delta}_{l}\partial_{z} & -\left(E_{-l}^{+}-\epsilon_{l}^{-}\partial_{z}^{2}\right)
\end{array}\right).
\end{eqnarray} 

As the signs of $\epsilon_{l}^{\pm}$ are the same, one can easily see that the $l=0$ Hamiltonian is essentially the same as a Kitaev chain. For $l \neq 0$, on the other hand, the Hamiltonian
does not describe a Kitaev chain. The PH symmetry is only present when both $\pm l$, besides the $\sigma=\pm$ sectors, are taken into account. In this 1D projection, the contributions
of the states in the whole radial direction are taken into account; in contrast, when probing the 3D system's LDOS at the center of the vortex, we filtered the contributions of $l=0$ and $l=1$ 
only. Because of this, PH symmetry is apparently broken in Figure \ref{fig:Peaks}. These considerations are more clearly seen by writing
\begin{eqnarray}
& & \left(\begin{array}{cc}
E_{l}^{+}-\epsilon_{l}^{+}\partial_{z}^{2} & -i\tilde{\Delta}_{l}\partial_{z}\\
-i\tilde{\Delta}_{l}\partial_{z} & E_{l}^{-}+\epsilon_{l}^{-}\partial_{z}^{2}
\end{array}\right)\\
&\approx& \left(\begin{array}{cc}
E_{l}^{+}-\epsilon_{l}\partial_{z}^{2} & -i\tilde{\Delta}_{l}\partial_{z}\\
-i\tilde{\Delta}_{l}\partial_{z} & -\left(E_{-l}^{+}-\epsilon_{l}\partial_{z}^{2}\right)
\end{array}\right)\\
& = & \frac{E_{l}^{+}-E_{-l}^{+}}{2}\rho_{0}+\rho_{z}\left(\frac{E_{l}^{+}+E_{-l}^{+}}{2}-\epsilon_{l}\partial_{z}^{2}\right) \nonumber\\
& &+\rho_{x}\left(-i\tilde{\Delta}_{l}\partial_{z}\right). \label{eq:proj}
\end{eqnarray}
The $\rho_0$ term does not vanish here (unless $l=0$), as usually happens. To see that indeed the system is PH symmetric, one has to take into account the full second quantized Hamiltonian
with all $\pm l$ pairs.

Likewise as above, the fluctuations may be written
\begin{eqnarray}
& &  \left(\begin{array}{cc}
\mathbf{R}\left(\tau\right)\cdot\mathbf{M}_{l,l^{'}}^{+} & 0\\ \label{eq:projfluc}
0 & \mathbf{R}\left(\tau\right)\cdot\mathbf{M}_{l,l^{'}}^{-}
\end{array}\right)=\\ 
&&\mathbf{R}\left(\tau\right)\cdot\left(\frac{\mathbf{M}_{l,l^{'}}^{+}+\mathbf{M}_{l,l^{'}}^{-}}{2}\right)\rho_{0}+\mathbf{R}\left(\tau\right)\cdot\left(\frac{\mathbf{M}_{l,l^{'}}^{+}-\mathbf{M}_{l,l^{'}}^{-}}{2}\right)\rho_{z}.\nonumber  
\end{eqnarray}
They couple diagonally in the $\sigma=\pm$ indices and also bring up an apparently PH breaking term.

This projected Hamiltonian is then equivalent to a p-wave wire network. This is a very unusual network as PH symmetry actually connects different wires while each wire has PH symmetry actually
broken. Although unusual, however, similar ideas have been considered in the literature \cite{gaidamauskas}. It is remarkable, in any case, that 
the 3D physics we started with ends up in such an exotic 1D scenario. 

\subsubsection{TPT signatures in 1D}

\begin{figure}[t]
\includegraphics[scale=0.35]{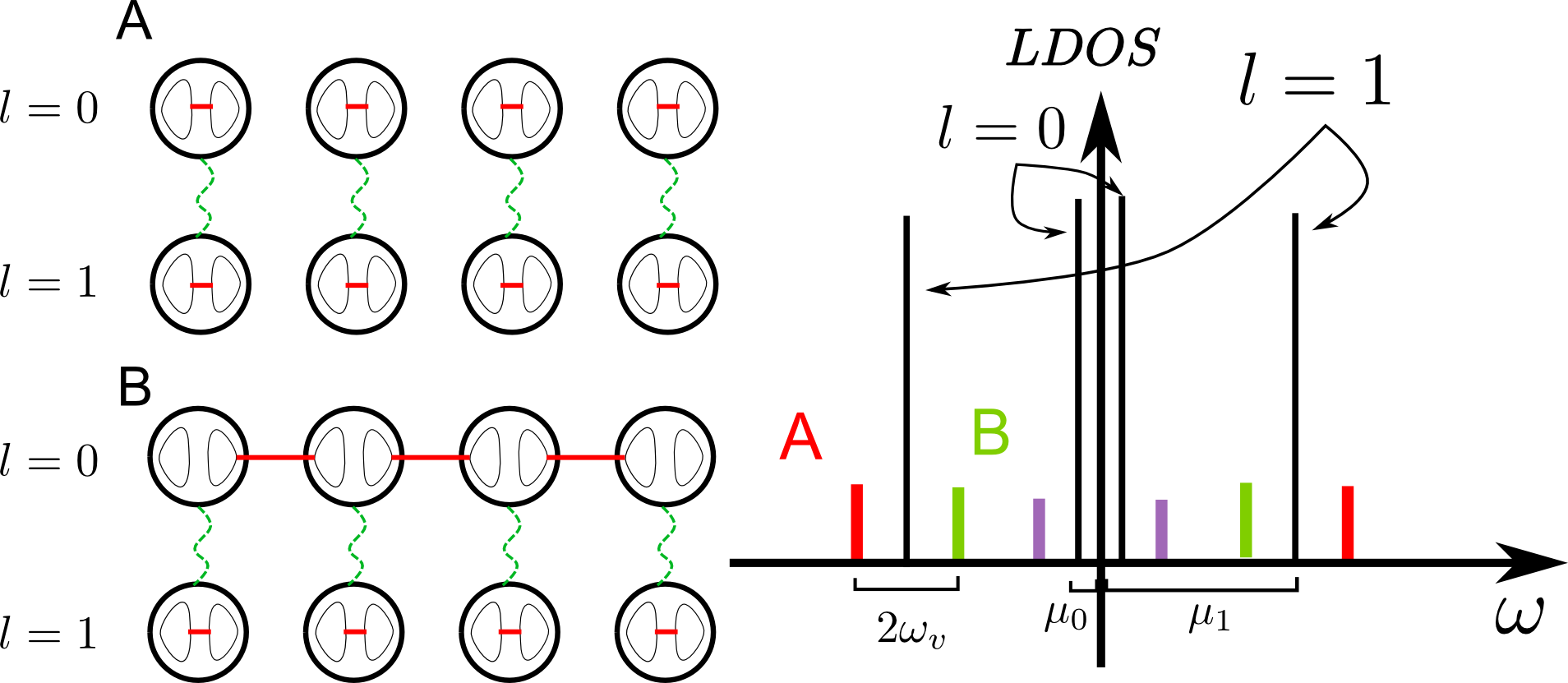}

\protect\caption{1D chain minimal model to study TPTs by quantum fluctuations. Large circles represent lattice complex fermions while the prolate circles represent the Majorana fermions inside. The $l=0$ represents a Kitaev chain which we drive through a topological phase transition while $l=1$ is kept at a large gap. Situations A and B correspond to the two deep topological regimes. The chains sites couplings are represented by green wriggly lines while Majorana hoppings are represented by the red lines. On the right, we display an schematic picture of the LDOS for this system. Large black peaks correspond to the peaks at the energy gaps $\pm  \mu_l$ with $k=0$, dispersion effects are not considered. As there is  single fluctuating coupling between the chains, only a single satellite peak appears besides each $\pm \mu_l$. The purple peaks correspond to the satellite peaks of $\pm\mu_0$. The read and green peaks differ by $2\omega_v$, the fluctuation energy scale, and correspond to the two topological 
regimes A and B on the left figure. Even if the satellite peaks jump at the same time, since the energy gap from wire $l=1$ is large, one may resolve the distinct situations with the sattelite peaks ``outside'' or ``inside'' the peaks from $\pm\mu_1$ as 
in the A or B situations, respectivelly. Notice that PH symmetry is explicitly respected in this context. \label{fig:1dchain}}

\end{figure}

From the above results, we may identify the minimal ingredients necessary to magnify the signatures of TPTs in 1D systems by a mechanism similar to as considered
in the vortex case. This minimal set of ingredients is undemanding. We list them in the context of Kitaev chains, as there seems to be a recent focus of interest in the literature concerning Kitaev
wires networks (see \cite{wakatsuki}, for example). We stress, however, that the same ingredients would suffice for other 1D topological systems, such as Su-Schrieffer-Heeger wires or Kitaev superconducting
\begin{enumerate}
\item A pair of gapped wires, one of which tunable through a TPT;
\item A diagonal fluctuating coupling between them;
\end{enumerate}
With these, one may reconstruct the important features of Eqs.\eqref{eq:proj} and \eqref{eq:projfluc}. This situation is illustrated in Figure \ref{fig:1dchain}.

Notice that the broken (PH) symmetry found in the projection on the vortex modes in the 3D scenario is not fundamental and is not included in our minimal list. It implies but a shift in the 
peaks in the spectrum, like as in the large peaks from $l=1$ in Fig. \ref{fig:Peaks}, and hence is unimportant. Also, a single pair of Kitaev chains is enough (the effects from $l=2$ and $l=-1$ in \eqref{eq:sigma0} and 
\eqref{eq:sigma1} are not important). This pair of wires could also be substituted by a single wire with a pair of low energy bands. The tight-binding
model for this is written
\begin{widetext}
 \begin{equation}
  H_{Kit}^{l}=-\mu_l\sum_{j}c_{j}^{l\dagger}c_{j}^{l}-\frac{1}{2}\sum_{j}\left(t_l c_{j}^{l\dagger}c_{j+1}^{l}+\Delta_l e^{i\phi_l}c_{j}^{l}c_{j+1}^{l}+H.c.\right),
 \end{equation}
\end{widetext}
where $\mu_l$ are the chemical potentials, $\Delta_l$ are the SC pairings, $\phi_l$ are the corresponding SC phases and $t_l$ the hopping amplitude for each wire. The index $l=0,\,1$ labels 
the two chains. Upon BdG doubling, it is easy to demonstrate that this reduces to \eqref{eq:proj} in k-space, without the $\rho_0$ term.

As for the fluctuation part of the Hamiltonian, one may have simply
\begin{equation}
 U=\sum_{j}c_{j}^{0\dagger}\Phi\left(\tau\right) c_{j}^{1}+H.c.,
\end{equation}
for a fluctuating coupling $\Phi$. This should lead to similar self-energy corrections to the wires energies as \eqref{eq:sigma0} and \eqref{eq:sigma1}, namely
\begin{equation}
\Sigma_{0,k}^{\sigma}\left(\omega\right) = \frac{A_{0;1}^{\sigma}}{\omega-sgn\left(E_{1,k}^{+;\sigma}\right)\omega_{v}-E_{1,k}^{\sigma}}
\end{equation}
and
\begin{equation}
\Sigma_{1,k}^{\sigma}\left(\omega\right) = \frac{A_{1;0}^{\sigma}}{\omega-sgn\left(E_{0,k}^{-;\sigma}\right)\omega_{v}-E_{0,k}^{\sigma}},
\end{equation}
where $\sigma$ gives the two Nambu components. The fluctuation frequency $\omega_v$ of $\Phi$ determes the new large energy scale. To find the positions of the peaks, one solves again
\begin{equation}
 \omega-E_{l,k}^{\sigma}-\Sigma_{l,k}^{\sigma}\left(\omega\right)=0
\end{equation}
which now leads to a single sattelite peak for each energy level. 

Importantly, the effects of the Magnus force are not necessary in the 1D case and, hence, a single fluctuating parameter is enough. 
This happens because one may (by ramping the chemical potential transversally to the wires, for example) keep a single wire well away from the phase transition with a 
large gap. Suppose, for example, wire $l=1$ is kept with a large gap. In this case, the sattelite peaks from the two sectors in this wire will stay always far away from each other. 
This way, by tuning the chemical potential from wire $l=0$, one can verify its phase transition by probing for the jumping in the satellite peaks of wire $l=1$.

As a final comment, out of the p-wave superconductivity context, one might work similarly with a set of Su-Schrieffer-Heeger (SSH) wires. In this case, the Hamiltonian will be similar to 
as the BdG Hamiltonian considered so far, with the caveat that the Nambu spinor now should be substituted by an ordinary spinor for a sublattice pseudo-spin degree of freedom. The gapping
parameters in this case will be given by staggered hopping amplitudes and chemical potentials. Formally, the problem is the same and one may extend the results discussed so far to this situation.

\section{Conclusions \label{sec:conc}}

Quantum fluctuations of vortex positions are ubiquitous and should manifest themselves at very low temperatures. We found out that, in the context of doped three dimensional topological insulators
these fluctuations may be exploited to magnify the signatures of topological vortex quantum phase transitions. This manifests at the LDOS at the vortex core by energy peaks which discontinuously
jump as function of the chemical potential. This finding also determined characteristic features of the low-energy Caroli-de Gennes-Matricon modes in this system which make them stand out as 
very distinct from standard s-wave Caroli-de Gennes modes, such as their spatial distribution and effects in the LDOS at the vortex core. Finally, our results also point to the possibility of capturing the effects of Magnus forces acting on the vortices, whose magnitude 
is directly related to the chemical potential values in which the topological phase transition induces peak position shifts.

The frequency of the position fluctuations plays an important role as it sets the scale of the peak jumps. In the context of high temperature superconductors, there are reports of this energy scale 
going up to meV\cite{bartoschomega}. It is important to point out that this frequency can be controlled to some extent and indeed 
increased depending on the properties of the vortex pinning potential. Recent developments in doping TIs with Niobium, which leads to the formation of magnetic moments in the bulk superconducting 
TI, can provide stronger pinning and so larger frequencies for the vortex fluctuation\cite{horghaemi}. Measured physical values of the vortex fluctuation frequencies and Magnus force frequency
in this system are not known to us at this point.

Cryogenic STM measurements are fundamental to uncover the discussed signatures. Situations with lighter and smaller vortices, whose zero-point motion effects would be stronger, could also be 
arranged as the vortex size is known to be strongly sensitive to temperature and magnetic field strength \cite{sonier}. For vortices of too minute sizes, however, the Taylor expansion method 
deployed here to derive the interactions  is not precise. In such cases, different approaches to the problem, such as used in \cite{Nikolic}, are necessary in order to obtain trustworthy predictions.
Also, a proper account for the effects of dispersion along the vortex may need detailed attention. It is beyond the scope of this work to consider these.

Finally, we studied the local physics along the vortex core. Projecting the Hamiltonian with the Caroli-de Gennes-Matricon wavefunctions we demonstrated explicitly that the vortex line
behaves as a Kitaev chain, with the corresponding topological phase transition. Further studying how the vortex position fluctuations are projected into this system allowed us to find some 
key ingredients which one may use to obtain new signatures of topological phase transitions in one-dimension.  A promising scenario lies in the study of the density of states 
upon fluctuations of the transversal coupling between a pair of neighboring gapped wires. Again, effects of dispersion along the wires still deserve attention. 

\section*{Acknowledgments}

The authors acknowledge insightful discussions with V. L. Quito, S. Sachdev, S. Gopalakrishnan, V. P. Nair, M. Sarachik and A. P. Polychronakos.  PLSL acknowledges support from FAPESP under grant 2009/18336-0.

\appendix
\section{Caroli-de Gennes modes \label{app:appA}} 

Here we present our numerical method to derive the spectrum of vortex modes for a stationary vortex in doped superconducting topological insulator \cite{pavan}, 
and compare the solutions with an analytical approximated ansatz. The latter will be used to study the effect of vortex quantum zero-point motion on the vortex spectrum.

The method which we apply is analogous to the one introduced by \cite{Gygi}
and \cite{Bartosch} in the context of ordinary superconductors. We
start by considering a cylinder space, infinite in the $z$ direction.
The Bogoliubov-de Gennes Hamiltonian with a static vortex centered at the origin reads
\begin{equation}
H_{BdG}^{0}=v_{D}\boldsymbol{\Gamma}\cdot\mathbf{P}-\mu\Sigma+\Gamma_{0}\left(m-\epsilon P^{2}\right)-\boldsymbol{\Lambda}\cdot\boldsymbol{\Delta}\left(\mathbf{r}\right), \label{eq:HBdG0}
\end{equation}
where the ''Dirac velocity'' $v_{D}$ is set to one throughout our
derivations and recovered to simplify the numerical calculations later.
The Dirac matrices obey $\left\{ \Gamma_{\mu},\Gamma_{\nu}\right\} =2\delta_{\mu\nu},\,\left\{ \Lambda_{a},\Lambda_{b}\right\} =2\delta_{ab}$
and $\left[\Gamma_{\mu},\Sigma\right]=\left\{ \Lambda_{a},\Sigma\right\} =0.$
Notice $\Sigma$ commutes with the kinetic Hamiltonian and is not
a ``mass'' term. In our basis, a choice for the representation follows

\begin{eqnarray}
\boldsymbol{\Gamma}=\rho_{z}\tau_{x}\boldsymbol{\sigma},\  &  &\  \boldsymbol{\Lambda}=\boldsymbol{\rho}\tau_{0}\sigma_{0}\\
\Gamma_{0}=\rho_{z}\tau_{z}\sigma_{0},\  &  &\  \Sigma=\rho_{z}\tau_{0}\sigma_{0}
\end{eqnarray}
with Nambu, orbital and spin spaces described by $\rho,$ $\tau$
and $\sigma$ Pauli matrices, respectively. Here $\boldsymbol{\Delta}\left(\mathbf{r}\right)=\Delta_{0}\left(r\right)\left(\cos\theta,\sin\theta\right)$
gives the pairing with a $\Delta_{0}\left(r\right)=\Delta_{0}\tanh\left(r/\xi\right)$
profile, to be concrete.

The vortex runs along the $z$ direction and translation invariance
allows us to consider the $k_{z}$ momentum; with the understanding
that only $k_{z}=0,\,\pi$ are topologically relevant, we take $k_{z}=0$
since we are looking only for the low energy Caroli-de Gennes-Matricon
modes \cite{pavan}.

The Hamiltonian commutes with the generalized angular momentum operator
$\tilde{L}_{z}=-i\partial_{\theta}-\frac{S_{z}+\Sigma}{2}$, where
$\Gamma_{x}\Gamma_{y}=i\rho_{0}\tau_{0}\sigma_{z}\equiv iS_{z}$.
This allows writing the solution spinors as
\begin{equation}
\chi_{l,n}\left(\mathbf{r}\right)=\frac{1}{\sqrt{2\pi}}e^{-i\left(l-\frac{S_{z}+\Sigma}{2}\right)\theta}\phi_{l,n}\left(r\right),
\end{equation}
where $l$ is an integer representing the standard angular momentum
and $n$ labels the many possible energies for a given $l$. At $k_{z}=0$,
the Hamiltonian obeys a further symmetry given by $\mathcal{M}=\rho_{0}\tau_{z}\sigma_{z}$.
Noticing that $\left\{ \mathcal{C},\mathcal{M}\right\} =0$ and naturally
$\left\{ \mathcal{C},H_{BdG}\right\} =0$, we see that the eigenvalues
of $\mathcal{M}$ also label particle and hole partners. This allows
one to separate $\phi_{l,n}\left(r\right)$ in four-spinors $\phi_{l,n}^{\pm}\left(r\right)$,
obeying corresponding Schrodinger's equations with projected Hamiltonians
$H^{\pm}$\cite{pavan},
\begin{equation}
H^{\pm}\phi_{l,n}^{\pm}=E_{l,n}^{\pm}\phi_{l,n}^{\pm}.
\end{equation}
We focus in $\phi_{l,n}^{+}$, noticing that $\phi_{l,n}^{-}=\mathcal{C}\phi_{-l,n}^{+}$
with $E_{l,n}^{-}=-E_{-l,n}^{+}$. The $4\times4$ reduced radial
Hamiltonian reads

\begin{eqnarray}
H^{+} & = & \rho_{z}\nu_{y}\left[-i\partial_{r}+i\nu_{z}\frac{1}{r}\left(l-\frac{\rho_{z}+\nu_{z}}{2}\right)\right]-\mu\rho_{z}-\Delta_{0}\left(r\right)\rho_{x}\nonumber \\
 &  & +\rho_{z}\nu_{z}\left[m+\epsilon\left(\frac{1}{r}\partial_{r}r\partial_{r}-\frac{1}{r^{2}}\left(l-\frac{\rho_{z}+\nu_{z}}{2}\right)^{2}\right)\right].
\end{eqnarray}
Here $\nu$ Pauli-matrices represent a spin-orbital coupled space.
Noticing that 
\begin{eqnarray}
a_{l} & = & \left(\partial_{r}+\frac{l}{r}\right)\\
a_{l}^{\dagger} & = & -\left(\partial_{r}-\frac{l-1}{r}\right)
\end{eqnarray}
act as operators which lower and raise the level of Bessel functions
(and $a_{l}^{\dagger}a_{l}$ gives the Bessel differential operator
itself), it is easy to find a proper basis to expand the states. If
$\Delta_{0}=0$, we recover a pair of topological insulator Hamiltonians
with spectra given by $E_{k}^{\pm\pm}=\pm\mu\pm\sqrt{k^{2}+m_{k}^{2}}$,
with $m_{k}=m-\epsilon k^{2}$ and $k$ a ``radial linear momentum''
quantum number. In the weak pairing approximation, since we are interested
only in the lowest energy modes, we solve for the eigenstates of the
TI Hamiltonian using the ladder operators above and project out the
bands from $E_{k}^{++}$ and $E_{k}^{--}$. Thus Fourier Bessel expanding
the radial wavefunctions as
\begin{equation}
\phi_{l,n}^{+}\approx\int dk\left(\begin{array}{c}
c_{l,k}^{n}f_{l,k}\left(r\right)\\
d_{l,k}^{n}g_{l,k}\left(r\right)
\end{array}\right),
\end{equation}
where
\begin{eqnarray}
f_{l,k}\left(r\right) & = & \frac{1}{\sqrt{\mathcal{N}_{k,l}^{+}}}\left(\begin{array}{c}
kJ_{l-1}\left(kr\right)\\
\left(m_{k}-\sqrt{m_{k}^{2}+k^{2}}\right)J_{l}\left(kr\right) \nonumber
\end{array}\right)\\
g_{l,k}\left(r\right) & = & \frac{1}{\sqrt{\mathcal{N}_{k,l}^{-}}}\left(\begin{array}{c}
\left(m_{k}+\sqrt{m_{k}^{2}+k^{2}}\right)J_{l}\left(kr\right)\\
-kJ_{l+1}\left(kr\right) \nonumber
\end{array}\right)\\
\end{eqnarray}
and $\mathcal{N}_{k,l}^{\pm}$ are normalization constants given by
$\mathcal{N}_{k}^{\pm}=2\left(k^{2}+m_{k}^{2}\mp m_{k}\sqrt{m_{k}^{2}+k^{2}}\right)\int_{0}^{\infty}rdrJ_{l}\left(kr\right)J_{l}\left(kr\right).$The
Schrodinger's equation reduces to

\begin{eqnarray}
\left(\begin{array}{cc}
T^{-} & \Delta^{+-}\\
\Delta^{+-T} & T^{+}
\end{array}\right)\Phi_{ln}^{+} & = & E_{l,n}^{+}\left(\mu\right)\Phi_{ln}^{+}\label{eq:EigenNum}
\end{eqnarray}
where
\begin{equation}
T_{k,k^{'}}^{\mp}=\left(\mp\mu\pm\sqrt{k^{2}+m_{k}^{2}}\right)\delta\left(k-k^{'}\right),
\end{equation}
with respective signs, 
\begin{equation}
\Delta_{l,k,k^{'}}^{+-}=\int rdrf_{l,k}^{T}\left(r\right)\left(\begin{array}{cc}
\Delta_{0}\left(r\right) & 0\\
0 & \Delta_{0}\left(r\right)
\end{array}\right)g_{l,k^{'}}\left(r\right)
\end{equation}
and the spinor is $\Phi_{ln}^{+}=\left(\left\{ c_{lk}^{n}\right\} ,\left\{ d_{lk}^{n}\right\} \right)^{T}.$

In terms of our original variables, the wavefunctions are then written
\begin{eqnarray}
\chi_{ln}^{+}\left(\mathbf{r}\right) & = & \left(\begin{array}{c}
\mathbf{u}_{ln}^{+}\left(\mathbf{r}\right)\\
\mathbf{v}_{ln}^{+}\left(\mathbf{r}\right)
\end{array}\right),\label{eq:stateuv}
\end{eqnarray}
where
\begin{eqnarray}
 &  & \mathbf{u}_{ln}^{+}\left(\mathbf{r}\right)=\nonumber \\
 &  & \int dk\frac{c_{lk}^{n}}{\sqrt{2\pi\mathcal{N}_{k}^{+}}}\left(\begin{array}{c}
e^{-i\left(l-1\right)\theta}kJ_{l-1}\left(kr\right)\\
0\\
0\\
e^{-il\theta}\left(m_{k}-\sqrt{m_{k}^{2}+k^{2}}\right)J_{l}\left(kr\right) \label{eq:ustate}
\end{array}\right)\nonumber \\
\\
 &  & \mathbf{v}_{ln}^{+}\left(\mathbf{r}\right)=\nonumber \\
 &  & \int dk\frac{d_{lk}^{n}}{\sqrt{2\pi\mathcal{N}_{k}^{-}}}\left(\begin{array}{c}
e^{-il\theta}\left(m_{k}+\sqrt{m_{k}^{2}+k^{2}}\right)J_{l}\left(kr\right)\\
0\\
0\\
ke^{-i\left(l+1\right)\theta}J_{l+1}\left(kr\right)
\end{array}\right)\nonumber \\
\end{eqnarray}
and the mirror (particle-hole) partners are built from $\chi_{ln}^{-}\left(\mathbf{r}\right)=\mathcal{C}\chi_{-ln}^{+}\left(\mathbf{r}\right)$. 

\begin{figure*}[t]
\begin{centering}
\includegraphics[scale=0.25]{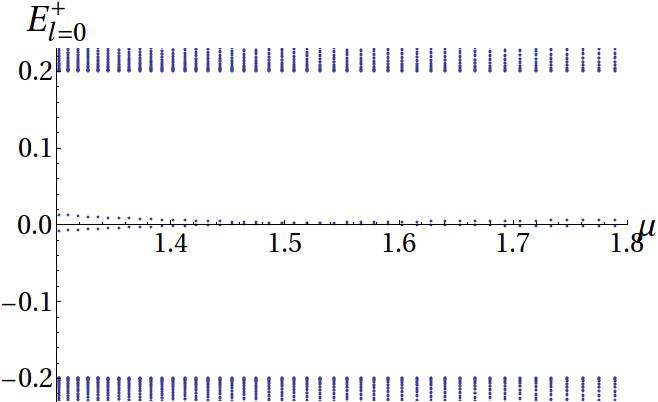}\includegraphics[scale=0.35]{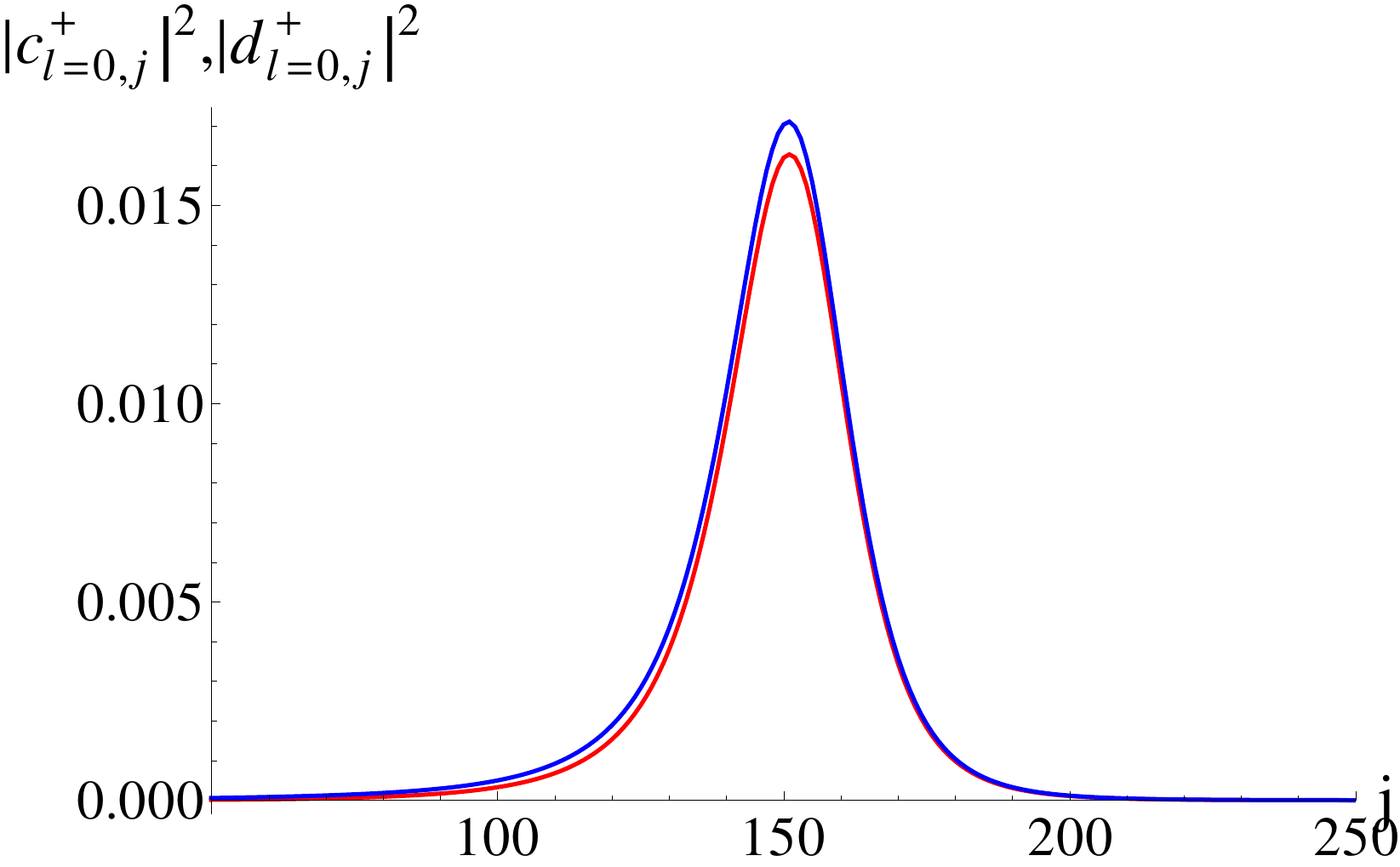}\includegraphics[scale=0.35]{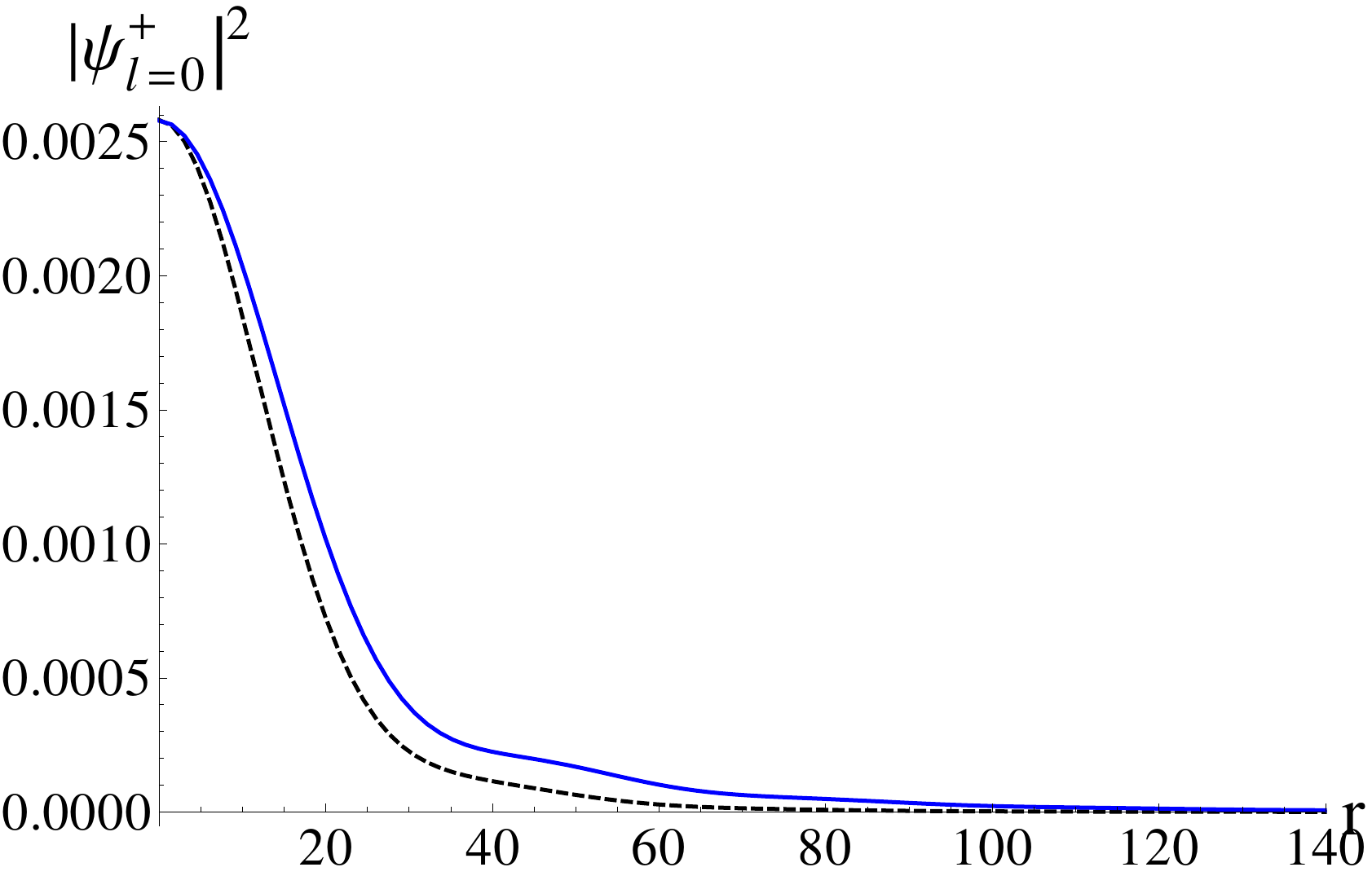}
\par\end{centering}

\protect\caption{Numerical results for the lowest CdG mode, $l=0$, in the $\sigma=+$
sector. Fixing the parameters demand some care as one needs to consider
a large enough region of k-space as to capture the TI band inversion
while considering Fermi energies large to guarantee $k_{F}\xi>1$
and at the same time close to the critical value $\mu_{C}\sim v_{D}\sqrt{m/\epsilon}$.
For all figures we use (disk size)$R=5000$, (number of modes) $N_{0}=300$,
$m=1$, $\epsilon=120$, $\Delta_{0}=0.2$, and use units with Dirac
velocity $v_{D}=15$. Thus $\xi=v_{D}/\pi\Delta_{0}\sim23$, and the
expected critical potential falls at $\mu_{C}\sim1.37$. (left) Energy
spectrum as function of the chemical potential. A clear gap is seen
at 0.2 with in gap modes. The two modes correspond to a vortex bound
mode, with positive slope, and a gapless expected edge mode, as can
be checked plotting the probability density in real space. (middle)
Momentum space distribution of the positive slope in-gap mode at chemical
potential close to the critical. The red and blue curves are associated
with $\left|c_{l=0,j}\right|^{2}$ and $\left|d_{l=0,j}\right|^{2}$
at discrete momenta $j\leftrightarrow k_{j}=\alpha_{l.j}/R$, respectively.
(right) Probability density in radial direction. The blue solid curve
corresponds to (\ref{eq:stateuv}) while the dashed line corresponds
to (\ref{eq:ApproxCdG}), demonstrating that our ansatz is indeed
a good approximation for the CdG modes wavefunctions.\label{fig:States}}
\end{figure*}

We then fix a finite radius $R$ for the cylinder size which forces
us to discretized $k\rightarrow\alpha_{l,j}/R$ where $\alpha_{l,j}$
are the $j$-th Bessel zeroes at each $l$ subspace. We fix a UV cutoff
at some (large) $N_{0}$-th Bessel zero. Diagonalizing the resulting
Hamiltonian leads to the spectrum shown in \ref{fig:States}. One
sees two in-gap modes, one corresponding to outer edge modes, which
we neglect, while the other corresponds to our desired vortex modes,
as can be checked by plotting their respective probability densities.

For the low energy states, $n\equiv n_{CdG}$ (a label which we drop
from now on), one may check that the spectrum follow the expected
\begin{equation}
E_{l}^{\pm}=\frac{\Delta^{2}}{E_{F}}\left(l\mp\frac{1}{2}\pm\frac{\phi\left(\mu\right)}{2\pi}\right)\label{eq:CdGEn}
\end{equation}
where $\phi\left(\mu\right)$ is the chemical potential dependent
Berry's phase. At the critical chemical potential $\phi\left(\mu_{C}\right)=\pi$.
It grows monotonically from $0$ to $2\pi$ with the chemical potential.
Noticing that the values of the momentum in $k$-space are strongly
localized at its Fermi value $k_{F}$, as one might expect, it is
easy to guess an analytical approximation for the wavefunctions which
satisfies their desired asymptotic behaviors (see \cite{pavan} for
details). We have
\begin{equation}
\chi^{+}=\mathcal{C}e^{-\frac{2}{v_{F}}\int_{0}^{r}dr^{'}\Delta\left(r^{'}\right)}\left(\begin{array}{c}
f\left(\theta,r\right)\\
g\left(\theta,r\right)
\end{array}\right),\label{eq:ApproxCdG}
\end{equation}
where
\begin{eqnarray}
 &  & f\left(\theta,r\right)=\nonumber \\
 &  & \frac{\mathcal{C}}{\sqrt{2\pi N_{k_{F}}^{+}}}\left(\begin{array}{c}
e^{-i\left(l-1\right)\theta}k_{F}J_{l-1}\left(k_{F}r\right)\\ 0
\\ 0
\\
e^{-il\theta}\left(m_{k_{F}}-\sqrt{m_{k_{F}}^{2}+k_{F}^{2}}\right)J_{l}\left(k_{F}r\right)
\end{array}\right)\nonumber \\
\\
 &  & g\left(\theta,r\right)=\nonumber \\
 &  & \frac{\mathcal{C}}{\sqrt{2\pi N_{k_{F}}^{-}}}\left(\begin{array}{c}
e^{-il\theta}\left(m_{k_{F}}+\sqrt{m_{k_{F}}^{2}+k_{F}^{2}}\right)J_{l}\left(k_{F}r\right)\\ 0
\\ 0
\\
e^{-i\left(l+1\right)\theta}k_{F}J_{l+1}\left(k_{F}r\right)
\end{array}\right)\nonumber \\
\end{eqnarray}
and the new normalizations read
\begin{equation}
N_{k_{F}}^{\pm}=2\left(k_{F}^{2}+m_{k_{F}}^{2}\mp m_{k_{F}}\sqrt{m_{k_{F}}^{2}+k_{F}^{2}}\right).
\end{equation}
Here, $\mathcal{C}$ is a normalization constant of order $\left(k_{F}/\xi\right)^{1/2}$.
In the main text we compare the analytical and numerical results for
the wavefunction at $l=0$ and $\mu\approx\mu_{c}$, showing that
the approximation indeed works.

\section{ Electronic Effective Interaction and Self-Energy \label{app:appB}} 

In this section we compute explicitly the electronic self-interaction
due to the interplay with the vortex fluctuations and the corresponding
self-energy in the GW approximation.

We start from the vortex effective action of the main text in frequency
space

\begin{eqnarray}
\mathcal{S}_{eff}^{vortex} & =\nonumber \\
 & = & \frac{m_{v}}{2}\int\frac{d\omega}{2\pi}\mathbf{R}^{\dagger}\left(i\omega\right)\left(\begin{array}{cc}
\omega^{2}+\omega_{0}^{2} & \omega_{c}\omega\\
-\omega_{c}\omega & \omega^{2}+\omega_{0}^{2} \nonumber
\end{array}\right)\mathbf{R}\left(i\omega\right) \\
\end{eqnarray}
and work at zero-temperature. Noticing that $\mathbf{R}^{\dagger}\left(i\omega\right)=\mathbf{R}\left(-i\omega\right)$,
we introduce a basis $R_{\pm}\left(i\omega\right)=\frac{R_{x}\left(i\omega\right)\pm iR_{y}\left(i\omega\right)}{\sqrt{2}}$
which diagonalizes the Lagrangian density as 
\begin{eqnarray}
\mathcal{S}_{eff}^{vortex} & = & \int\frac{d\omega}{2\pi}\left(R_{-}^{\dagger}\left(i\omega\right),\, R^{\dagger}_{+}\left(i\omega\right)\right)\mathbb{D}_{0}^{v}\left(i\omega\right)^{-1}\left(\begin{array}{c}
R_{-}\left(i\omega\right)\\
R_{+}\left(i\omega\right)
\end{array}\right),\nonumber \\
\end{eqnarray}
with
\begin{equation}
\mathbb{D}_{0}^{v}\left(i\omega\right)=\left(\begin{array}{cc}
D_{-}^{-1} & 0\\
0 & D_{+}^{-1}
\end{array}\right)
\end{equation}
and the Green's functions $D_{\mp}\left(i\omega\right)=\frac{m_{v}}{2}\left(\left(\omega\pm i\omega_{c}/2\right)^{2}+\omega_{v}^{2}\right)$.
This sets the two important energy scales dictated by the vortex fluctuations
as $\omega_{c}$, from the Magnus force, and $\omega_{v}=\sqrt{\omega_{0}^{2}+\omega_{c}^{2}/4}$,
from the harmonic trap. 

As discussed in the former section, the low-energy modes divide into
two Hilbert space sectors related by a $z$-mirror/particle-hole symmetry.
Each sector is subject to an effective potential arising after the
integration of the vortex 0D field theory. From equations \ref{eq:Svrtx} we may write:

\begin{widetext}
\begin{equation}
e^{-V_{eff}^{\sigma}\left[\bar{\psi}_{l}^{\sigma},\psi_{l}^{\sigma}\right]}\propto\int\mathcal{D}\left[\mathbf{R}\right]e^{-S_{eff}^{vortex}+\int d\tau\sum_{l,l^{'}}\mathbf{R}\left(\tau\right)\cdot\mathbf{M}_{l,l^{'}}^{\sigma}\bar{\psi}_{l}^{\sigma}\left(\tau\right)\psi_{l^{'}}^{\sigma}\left(\tau\right)}.
\end{equation}
Define $\mathbf{U}^{\sigma}\left(i\omega\right)=\sum_{l,l^{'}}\int\frac{d\nu}{2\pi}\mathbf{M}_{l,l^{'}}^{\sigma}\bar{\psi}_{l}^{\sigma}\left(iv+i\omega\right)\psi_{l^{'}}^{\sigma}\left(iv\right)$ and rewrite the scalar products in terms of the $R_{\pm}\left(i\omega\right)$
coordinates and $M^{\alpha;\sigma}=\frac{1}{2}\left(M_{x}+\alpha iM_{y}\right)$, with $\alpha=\pm$. Then

\begin{equation}
M_{l,l^{'}}^{+;\sigma}=\left(M_{l^{'},l}^{-;\sigma}\right)^{*}=\int d^{2}r\left[\mathbf{u}_{l}^{\sigma}\left(\mathbf{r}\right)^{\dagger}\partial_{\bar{z}}\Delta\mathbf{v}_{l^{'}}^{\sigma}\left(\mathbf{r}\right)+\mathbf{v}_{l}^{\sigma}\left(\mathbf{r}\right)^{\dagger}\partial_{\bar{z}}\Delta^{\dagger}\mathbf{u}_{l^{'}}^{\sigma}\left(\mathbf{r}\right)\right].\label{eq:Overlap}
\end{equation}

 This allows, with a careful consideration of
positive and negative frequencies, integration over the vortex degrees
of freedom, leading to the effective action of the electronic modes
as
\begin{eqnarray}
S_{eff}^{\sigma}\left[\bar{\psi}^{\sigma},\psi^{\sigma}\right] & = & \sum_{l}\int\frac{d\tilde{\omega}}{2\pi}\bar{\psi}_{l}^{\sigma}\left(i\tilde{\omega}\right)\left(i\tilde{\omega}-E_{l}^{\sigma}\right)\psi_{l}^{\sigma}\left(i\tilde{\omega}\right)\\
 &  & -\int\frac{d\tilde{\omega}}{2\pi}\left[\frac{1}{4}\left(U_{-}^{\sigma\dagger}D_{+}^{-1}U_{-}^{\sigma}+U_{+}^{\sigma\dagger}D_{-}^{-1}U_{+}^{\sigma}\right)\right],\nonumber 
\end{eqnarray}
where $U_{\alpha}^{\sigma}=\frac{1}{2}\left(U_{x}^{\sigma}+\alpha iU_{y}^{\sigma}\right)$.
A tedious but straightforward simplification leads to the effective
electronic self-interaction
\begin{eqnarray}
V_{eff}^{\sigma}\left[\bar{\psi}_{l}^{\sigma},\psi_{l}^{\sigma}\right] & = & \frac{1}{2}\sum_{l,l^{'},n,n^{'}}\int\frac{d\tilde{\omega}}{2\pi}\int\frac{d\tilde{\nu}}{2\pi}\int\frac{d\tilde{\nu}^{'}}{2\pi}\times\nonumber \\
 &  & \bar{\psi}_{l}^{\sigma}\left(i\tilde{\nu}+i\tilde{\omega}\right)\bar{\psi}_{n}^{\sigma}\left(i\tilde{\nu}^{'}-i\tilde{\omega}\right)V_{l,l^{'},n,n^{'}}^{\sigma}\left(i\tilde{\omega}\right)\psi_{l^{'}}^{\sigma}\left(i\tilde{\nu}\right)\psi_{n^{'}}^{\sigma}\left(i\tilde{\nu}^{'}\right)
\end{eqnarray}
where 
\begin{equation}
V_{l,l^{'},n,n^{'}}^{\sigma}\left(i\tilde{\omega}\right)=-\frac{1}{m_{v}}\sum_{\alpha=\pm}\left[\frac{\left(M_{l,l^{'}}^{\alpha;\sigma}\right)^{\dagger}M_{n,n^{'}}^{\alpha;\sigma}}{\left(\left(\tilde{\omega}+\alpha i\omega_{c}/2\right)^{2}+\omega_{v}^{2}\right)}\right].\label{eq:effpot}
\end{equation}
\end{widetext}From (\ref{eq:stateuv}), the matrix elements have
a simple form
\begin{equation}
M_{l,l^{'}}^{\alpha;\sigma}=\int d^{2}r\left[\mathbf{u}_{m}^{\sigma}\left(\mathbf{r}\right)^{\dagger}\partial_{\bar{z}}\Delta\mathbf{v}_{m^{'}}^{\sigma}\left(\mathbf{r}\right)+\mathbf{v}_{m}^{\sigma}\left(\mathbf{r}\right)^{\dagger}\partial_{\bar{z}}\Delta^{\dagger}\mathbf{u}_{m^{'}}^{\sigma}\left(\mathbf{r}\right)\right]\label{eq:matelem}
\end{equation}
which also shows the convenient fact that $M_{l,l^{'}}^{+;\sigma}=M_{l^{'},l}^{-;\sigma*}$.

Interaction (\ref{eq:effpot}) shows a screened Coulomb-like retarded
interaction. The self-energy in the GW approximation comes now from
a simple 1-loop calculation
\begin{eqnarray}
\Sigma_{l}^{\sigma}\left(i\tilde{\omega}\right) & = & -\sum_{l^{'}}V_{l,l,l^{'},l^{'}}^{\sigma}\left(0\right)\int_{\omega}G_{l}^{0\sigma}\left(i\omega\right)\nonumber \\
 &  & +\sum_{l^{'}}\int_{\omega}V_{l,l^{'},l^{'},l}^{\sigma}\left(i\tilde{\omega}-i\omega\right)G_{l^{'}}^{0\sigma}\left(i\omega\right).
\end{eqnarray}
The first term vanishes. The second must be considered with care as
the pole structure is sensitive to the structure of the energy levels.
An integration over the complex plane gives the self-energy of the
main text
\begin{equation}
\Sigma_{l}^{\sigma}\left(i\tilde{\omega}\right)=\sum_{l^{'}}\sum_{\alpha=\pm}\frac{A_{l;l^{'}}^{\alpha;\sigma}}{\left(i\tilde{\omega}-\left(sgn\left(\Xi_{l^{'}}^{\alpha;\sigma}\right)\omega_{v}+E_{l^{'}}^{\sigma}\right)-\alpha\omega_{c}/2\right)},
\end{equation}
where $A_{l;l^{'}}^{\alpha;\sigma}\equiv\frac{\left|M_{l,l^{'}}^{\alpha;\sigma}\right|^{2}}{m_{v}\omega_{v}}$
and $\Xi_{l^{'}}^{\alpha;\sigma}\equiv E_{l^{'}}^{\sigma}+\alpha\omega_{c}/2$.

To calculate the matrix elements one may make use of the Feynman-Hellman
relations, adapted to our Hamiltonian and in a finite cylinder. A
long calculation making full use of Bessel function relations gives
finally
\begin{eqnarray}
M_{l,l^{'}}^{+;+} & = & \frac{\delta_{l^{'},l+1}}{2}\sum_{j,j^{'}}c_{lj}\left[\left(E_{l+1}^{+}-E_{l}^{+}\right)\mathcal{K}_{j,j^{'}}^{l+}-\mathcal{L}_{j,j^{'}}^{l+}\right]c_{l+1j^{'}}\nonumber \\
 &  & +\frac{1}{2}\sum_{j,j^{'}}d_{lj}\left[\left(E_{l+1}^{+}-E_{l}^{+}\right)\mathcal{K}_{j,j^{'}}^{l-}-\mathcal{L}_{j,j^{'}}^{l-}\right]d_{l+1j^{'}}\nonumber \\
\end{eqnarray}
with\begin{widetext}

\begin{equation}
\mathcal{K}_{j,j^{'}}^{l\pm}=\mbox{sgn}\left(l+1/2\right)\left(-1\right)^{j-j^{'}}\frac{\alpha_{jl}\alpha_{j^{'}l+1}}{R\left(\alpha_{j^{'}l+1}^{2}-\alpha_{jl}^{2}\right)}\frac{\mathcal{M}_{jl}^{\pm}\mathcal{M}_{j^{'}l+1}^{\pm}+\left(\frac{\alpha_{j^{'}l+1}}{R}\right)^{2}}{\sqrt{\left(\left(\frac{\alpha_{jl}}{R}\right)^{2}+\mathcal{M}_{jl}^{\pm}\right)\left(\left(\frac{\alpha_{j^{'}l+1}}{R}\right)^{2}+\mathcal{M}_{j^{'}l+1}^{\pm}\right)}},
\end{equation}
and
\begin{equation}
\mathcal{L}_{j,j^{'}}^{l\pm}=\mbox{sgn}\left(l+1/2\right)\left(-1\right)^{j-j^{'}}\frac{2\epsilon\alpha_{jl}\alpha_{j^{'}l+1}}{R^{3}}\frac{\left(\frac{\left(l\mp1\right)\left(l+1\mp1\right)}{R^{2}}+\mathcal{M}_{jl}^{\pm}\mathcal{M}_{j^{'}l^{'}}^{\pm}\right)}{\sqrt{\left(\left(\frac{\alpha_{jl}}{R}\right)^{2}+\mathcal{M}_{jl}^{\pm}\right)\left(\left(\frac{\alpha_{j^{'}l+1}}{R}\right)^{2}+\mathcal{M}_{j^{'}l+1}^{\pm}\right)}}.
\end{equation}
\end{widetext}Here, $R$ is the cylinder finite radius, $\alpha_{jl}$
is the j-th zero of the l-th Bessel function and $\mathcal{M}_{jl}^{\pm}=m_{j,l}^{2}\mp m_{j,l}\sqrt{m_{j,l}^{2}+\left(\frac{\alpha_{jl}}{R}\right)^{2}}$
with $m_{j,l}=m-\epsilon\alpha_{jl}/R$. The other matrix elements
may be found from

\begin{eqnarray}
M_{l,l^{'}}^{\alpha;-} & = & -M_{-l^{'},-l}^{\alpha;+}\\
M_{l,l^{'}}^{-;\sigma} & = & \left(M_{l^{'},l}^{+;\sigma}\right)^{*}\equiv\left(M_{l,l^{'}}^{+;\sigma}\right)^{\dagger}.
\end{eqnarray}
These expressions are very similar to Bartosch's, corrected for spin-orbit
coupled states.

\section{ Peak Analysis \label{app:appC}}

Here we describe in detail the determination the relative sizes and positions
of the tunneling conductance peaks. We start rewriting,
\begin{eqnarray}
\rho\left(\mathbf{r},\omega\right) & = & \sum_{\sigma=\pm}\rho_{\sigma}\left(\mathbf{r},\omega\right)\\
\rho_{\sigma}\left(\mathbf{r},\omega\right) & = & -\frac{1}{\pi}Im\sum_{l}\frac{\left|\mathbf{u}_{l}^{\sigma}\left(\mathbf{r}\right)\right|^{2}}{\omega-E_{l}^{\sigma}-\Sigma_{l}^{\sigma}+i\epsilon},
\end{eqnarray}
using the vortex-modes eigenbasis. STM measurements probe the tunneling conductance
\begin{equation}
G\left(\mathbf{r},\omega\right)=-\frac{G_{0}}{\rho_{0}}\int d\omega^{'}\rho\left(\mathbf{r},\omega+\omega^{'}\right)f^{'}\left(\omega^{'}\right), \label{eq:Gcond}
\end{equation}
where $f\left(\omega\right)$is the Fermi distribution. 

At zero-temperature this reduces simply to the LDOS, up to a constant.
At finite temperature we may write
\begin{eqnarray}
G\left(\mathbf{r},\omega\right)/G_{0} & = & -\frac{1}{\rho_{0}}\sum_{l,\sigma=\pm}\sum_{i}\frac{\left|u_{l}^{\sigma}\left(\mathbf{r}\right)\right|^{2}}{\left|1-\frac{\partial\Sigma_{l}^{\sigma}\left(\omega_{l,\sigma,0}^{i}\right)}{\partial\omega}\right|}\label{eq:LDOS}\\
 &  & \times f^{'}\left(\omega_{l,\sigma,0}^{i}-\omega\right),
\end{eqnarray}
where $\omega_{l,\sigma,0}^{i}$ is the $i$-th solution to

\begin{equation}
\omega-E_{l}^{\sigma}-\Sigma_{l}^{\sigma}\left(\omega\right)=0.\label{eq:peakposition}
\end{equation}
This represents a cubic equation, thus with three solutions. While (\ref{eq:peakposition}) determines where are the
relative positions of the peaks in energy space, the derivatives $\frac{\partial\Sigma_{l}^{\sigma}\left(\omega_{l,\sigma,0}^{i}\right)}{\partial\omega}$
will fix the peaks relative sizes.

We focus most of our analysis at $\left|\mathbf{r}\right|=0$, which, from (\ref{eq:stateuv}), means that only the states with $l=0,\,1$
give non-vanishing contributions. The relevan self-energy contributions were considered in the main text in equations \eqref{eq:sigma0} and \eqref{eq:sigma1}.
To determine the relative sizes and positions of the peaks, we examine
the derivatives of the self-energy, as well as equation (\ref{eq:peakposition})
explicitly.

\subsection{Peak sizes}

The derivatives of the self-energies read, after some simplification

\begin{widetext}

\begin{eqnarray}
\frac{d\Sigma_{1}^{\sigma}\left(\omega\right)}{d\omega} & = & -\frac{A_{1;2}^{+;\sigma}}{\left(\Delta\omega_{1}^{\sigma}-\delta-\omega_{c}/2-\omega_{v}\right)^{2}}-\frac{A_{0;1}^{+;\sigma}}{\left(\Delta\omega_{1}^{\sigma}+\delta+\omega_{c}/2-\sigma sgn\left(\mu-\bar{\mu}_{\sigma}\right)\omega_{v}\right)^{2}}\\
\frac{d\Sigma_{0}^{\sigma}\left(\omega\right)}{d\omega} & = & -\frac{A_{0;1}^{+;\sigma}}{\left(\Delta\omega_{0}^{\sigma}-\delta-\omega_{c}/2-\omega_{v}\right)^{2}}-\frac{A_{-1;0}^{+;\sigma}}{\left(\Delta\omega_{0}^{\sigma}+\delta+\omega_{c}/2+\omega_{v}\right)^{2}},
\end{eqnarray}
\end{widetext}where $\Delta\omega_{l}^{\sigma}=\omega-E_{l}^{\sigma}$ and $\delta$ is the mini-gap.

The matrix elements are much smaller than the other physical quantities.
Dimensional analysis and explicit manipulation of (\ref{eq:Overlap})
shows that, at constant $\omega_{v}/\Delta_{0}$, these overlaps sizes
depend on the coherence length as $\xi^{-5}$\cite{Bartosch}. The
peak sizes, nevertheless, are going to be sensitive to $A_{l;l^{'}}^{\alpha;\sigma}$.
As will be seen in the next subsection, the satellite peaks positions
are dominated by the vortex oscillation frequency $\omega_{v}$. Plugging
$\Delta\omega_{l}^{\sigma}\approx0$ or $\Delta\omega_{l}^{\sigma}\approx\pm\omega_{v}$
one sees that $d\Sigma_{l}^{\sigma}\left(\omega\right)/d\omega$ is
small (concretely it is$\propto A_{l,l^{'}}^{+;\sigma}/\omega_{v}^{2}\ll1$)
at $\Delta\omega_{0}^{+}\approx0$ while it may be larger at $\Delta\omega_{0}^{\sigma}\approx\pm\omega_{v}$,
going as $\sim-A_{0;1}^{+;+}\left[\frac{1}{s^{2}}\right]$, where
$s=\frac{\delta+\omega_{c}/2}{2\omega_{v}}$. The latter case reduces greatly the size
of the satellite peaks from $l=0$, similarly as pointed by Bartosch
et al.\cite{Bartosch}.

\subsection{peak positions}

Our last goal is to explain the positions of the peaks as function
of the chemical potential, demonstrating that they are much less sensitive
to the matrix elements than the peak sizes and that they are mainly
fixed by the vortex fluctuation frequency, which might be much larger
than the other energy scales of the problem.


 Simplifying the self-energy and plugging into (\ref{eq:peakposition}),
shows that independent of chemical potential, for $l=0$ we have
\begin{equation}\begin{split}
\Delta\omega_{0}^{\sigma} & \left[\left(\Delta\omega_{0}^{\sigma}\right)^{2}-\left(\delta+\omega_{c}/2+\omega_{v}\right)^{2}+\left(A_{0;1}^{\sigma;+}+A_{-1;0}^{\sigma;+}\right)\right] \\
& +\left(\omega_{v}+\delta+\omega_{c}/2\right)\left(A_{0;1}^{\sigma;+}-A_{-1;0}^{\sigma;+}\right)  =  0.
\end{split}\end{equation}
Using $A_{0;1}^{\sigma;+}\approx A_{-1;0}^{\sigma;+}$  we get  results similar to reference \cite{Bartosch} for ordinary $s$-wave superconductor. Since the matrix elements
are much smaller than the other parameters, we can neglect them in above equation. We then get

\begin{equation}
\Delta\omega_{0}^{\sigma}\left[\left(\Delta\omega_{0}^{\sigma}\right)^{2}-\left(\delta+\omega_{c}/2+\omega_{v}\right)^{2}\right]=0,
\end{equation}
for any $\mu$. This gives a central and two satellite peaks at, respectively

\begin{eqnarray}
\Delta\omega_{0}^{\sigma} & = & 0\\
\Delta\omega_{0}^{\sigma} & = & \left(\omega_{v}+\delta+\omega_{c}/2\right)\\
\Delta\omega_{0}^{\sigma} & = & -\left(\omega_{v}+\delta+\omega_{c}/2\right).
\end{eqnarray}

For $l=1$, we may as well neglect the contributions from the matrix
elements. For $\mu<\bar{\mu}_-$,
\begin{eqnarray}
\Delta\omega_{1}^{-}\left[\left(\Delta\omega_{1}^{-}-\omega_{v}\right)^{2}-\left(\delta+\omega_{c}/2\right)^{2}\right]  &= & 0\\
\Delta\omega_{1}^{+}\left[\left(\Delta\omega_{1}^{+}\right)^{2}-\left(\omega_{v}+\delta+\omega_{c}/2\right)^{2}\right] & = & 0.
\end{eqnarray}
So we have peaks at 

\begin{eqnarray}
\Delta\omega_{1}^{-} & = & 0\\
\Delta\omega_{1}^{-} & = & \omega_{v}+\left(\delta+\omega_{c}/2\right)\\
\Delta\omega_{1}^{-} & = & \omega_{v}-\left(\delta+\omega_{c}/2\right)
\end{eqnarray}

and

\begin{eqnarray}
\Delta\omega_{1}^{+} & = & 0\\
\Delta\omega_{1}^{+} & = & \left(\omega_{v}+\delta+\omega_{c}/2\right)\\
\Delta\omega_{1}^{+} & = & -\left(\omega_{v}+\delta+\omega_{c}/2\right).
\end{eqnarray}

For $\mu>\bar{\mu}_+$ the role of $+$ and $-$ in above equations are exchanged. Since the total density is the sum of contributions form both $\sigma=\pm$ sectors, and the gap between
$E_{1}^{+}$ and $E_{1}^{-}$ goes as $\delta (1/2-\phi/2\pi)$, the LDOS in the two regimes of  $\mu<\bar{\mu}_-$ and  $\mu>\bar{\mu}_+$ look the same.

We now get to the most important regime of  $\bar{\mu}_-<\mu<\bar{\mu}_+$. The position of the peaks for both $\sigma=\pm$ sectors are at

\begin{eqnarray}
\Delta\omega_{1}^{\sigma} & = & 0\\
\Delta\omega_{1}^{\sigma} & = & \left(\omega_{v}+\delta+\omega_{c}/2\right)\\
\Delta\omega_{1}^{\sigma} & = & -\left(\omega_{v}+\delta+\omega_{c}/2\right).
\end{eqnarray}

Clearly, as $\mu$ crossed $\bar{\mu}_-$ the third peak for $\sigma=-$ sector is shifted by $-2\omega_v$ and this leads to a clear modification of LDOS which persists up the $\mu=\bar{\mu}_+$ at which the peak form the $\sigma=+$ sector moves by $2\omega_v$ 
and recovers the original LDOS. 

The ``creation'' of a satellite peak at positive energy should not
happen without an accompanying compensation of a positive energy peak
jumping into negative energies. Indeed, such a compensation does occur
for the contribution of $l=-1$ (which exchanging angular momentum
with the vortex motion is connected to $l=-2$ and $l=0$, the latter
giving the jump.) It just turns out that, since the spatial dependence
of the LDOS is determined by $u_{l}^{\sigma}\left(\mathbf{r}\right)$,
as can be seen from (\ref{eq:LDOS}), the peaks from $l=-1$ do not
contribute to the LDOS at the center of the vortex, $r=0$. The peaks
from $l=-1$ should contribute to the LDOS at a distance $\sim k_{F}^{-1}$
from the vortex center, which should be of the order of ten Angstroms
in a superconducting TI. This can be resolved with the current STM
technology.

\bibliographystyle{apsrev4-1}
\end{document}